\documentclass[pra,superscriptaddress,twocolumn,longbibliography]{revtex4-1}
\usepackage{amsmath,amssymb}
\usepackage{subfigure,bbm,times}
\usepackage[T1]{fontenc}
\usepackage{braket}
\usepackage{graphicx}
\usepackage{float}
\usepackage{verbatim}

\usepackage{array,amsfonts}
\usepackage{dcolumn,tabularx,booktabs}
\usepackage{bm}
\usepackage{color}
\usepackage[colorlinks,citecolor=blue]{hyperref}

\begin{document}
\hyphenpenalty=5000
\tolerance=1000

\title{Experimental control of remote spatial indistinguishability of photons to realize entanglement and teleportation}

\author{Kai Sun}
\affiliation{CAS Key Laboratory of Quantum Information, University of Science and Technology of China, Hefei 230026, People's Republic of China}
\affiliation{Synergetic Innovation Center of Quantum Information and Quantum Physics, University of Science and Technology of China, Hefei 230026, People's Republic of China}

\author{Yan Wang}
\affiliation{CAS Key Laboratory of Quantum Information, University of Science and Technology of China, Hefei 230026, People's Republic of China}
\affiliation{Synergetic Innovation Center of Quantum Information and Quantum Physics, University of Science and Technology of China, Hefei 230026, People's Republic of China}

\author{Zheng-Hao Liu}
\affiliation{CAS Key Laboratory of Quantum Information, University of Science and Technology of China, Hefei 230026, People's Republic of China}
\affiliation{Synergetic Innovation Center of Quantum Information and Quantum Physics, University of Science and Technology of China, Hefei 230026, People's Republic of China}

\author{Xiao-Ye Xu}
\affiliation{CAS Key Laboratory of Quantum Information, University of Science and Technology of China, Hefei 230026, People's Republic of China}
\affiliation{Synergetic Innovation Center of Quantum Information and Quantum Physics, University of Science and Technology of China, Hefei 230026, People's Republic of China}

\author{Jin-Shi Xu}
\email{jsxu@ustc.edu.cn}
\affiliation{CAS Key Laboratory of Quantum Information, University of Science and Technology of China, Hefei 230026, People's Republic of China}
\affiliation{Synergetic Innovation Center of Quantum Information and Quantum Physics, University of Science and Technology of China, Hefei 230026, People's Republic of China}

\author{Chuan-Feng Li}
\email{cfli@ustc.edu.cn}
\affiliation{CAS Key Laboratory of Quantum Information, University of Science and Technology of China, Hefei 230026, People's Republic of China}
\affiliation{Synergetic Innovation Center of Quantum Information and Quantum Physics, University of Science and Technology of China, Hefei 230026, People's Republic of China}

\author{Guang-Can Guo}
\affiliation{CAS Key Laboratory of Quantum Information, University of Science and Technology of China, Hefei 230026, People's Republic of China}
\affiliation{Synergetic Innovation Center of Quantum Information and Quantum Physics, University of Science and Technology of China, Hefei 230026, People's Republic of China}

\author{Alessia Castellini}
\affiliation{Dipartimento di Fisica e Chimica - Emilio Segr\`e, Universit\`a di Palermo, via Archirafi 36, 90123 Palermo, Italy}

\author{Farzam Nosrati}
\affiliation{Dipartimento di Ingegneria, Universit\`{a} di Palermo, Viale delle Scienze, Edificio 9, 90128 Palermo, Italy}
\affiliation{INRS-EMT, 1650 Boulevard Lionel-Boulet, Varennes, Qu\'{e}bec J3X 1S2, Canada}

\author{Giuseppe Compagno}
\affiliation{Dipartimento di Fisica e Chimica - Emilio Segr\`e, Universit\`a di Palermo, via Archirafi 36, 90123 Palermo, Italy}

\author{Rosario Lo Franco}
\email{rosario.lofranco@unipa.it}
\affiliation{Dipartimento di Fisica e Chimica - Emilio Segr\`e, Universit\`a di Palermo, via Archirafi 36, 90123 Palermo, Italy}
\affiliation{Dipartimento di Ingegneria, Universit\`{a} di Palermo, Viale delle Scienze, Edificio 6, 90128 Palermo, Italy}

\begin{abstract}

Remote spatial indistinguishability of identical subsystems as a direct controllable quantum resource at distant sites has not been yet experimentally proven.
We design a setup capable to tune the spatial indistinguishability of two photons by independently adjusting their spatial distribution in two distant regions, which leads to polarization entanglement starting from uncorrelated photons.
The amount of entanglement uniquely depends on the degree of remote spatial indistinguishability, quantified by an entropic measure $\mathcal{I}$, which enables teleportation with fidelities above the classical threshold. 
This experiment realizes a basic nonlocal entangling gate by the inherent nature of identical quantum subsystems.

\end{abstract}

\date{\today }

\maketitle

Discovering how fundamental properties of quantum constituents can facilitate preparation and control of composite systems is strategic for the scientific progress. In fact, this achievement has an impact both on the advance of our knowledge of the basic features of the natural world and on the development of quantum-enhanced technologies, simplifying the experimental challenges. Many-body quantum networks are usually composed of identical building blocks (subsystems or particles of the same species), such as atoms, electrons, photons or generic qubits \cite{laddReview,exptenphoton,fivequbitexp,RevModPhys.81.1051,benatti2013PRA,QEMreview}. Indistinguishability of identical subsystems thus emerges as an inherent quantum feature that may play a role in quantum information processing.

The usual requirement to implement quantum tasks in many-body systems is that the qubits are individually addressed \cite{vedralReview,horo2009}. For nonidentical qubits, this requirement is fulfilled by local operations and classical communication (LOCC), where the term ``local'' refers to particle-locality independently of their spatial configuration \cite{horo2009}. On the other hand, identical qubits are not in general individually addressable \cite{tichy2011essential}, spatial distribution of their wave functions becoming crucial. Despite the long debate concerning formal aspects on entanglement of identical particles \cite{ghirardi2004general,benatti2014review,wiseman2003PRL,Li2001PRA,
Paskauskas2001PRA,cirac2001PRA,zanardiPRA,eckert2002AnnPhys,balachandranPRL,
sasaki2011PRA,giulianoEPJD,buscemi2007PRA,bose2002indisting,bose2013,tichyFort,
plenio2014PRL,LFCSciRep,sciaraSchmidt,compagno2018dealing,morris2019} and some proposals using particle identity for quantum protocols \cite{benatti2014NJP,QEMreview,omarIJQI,PhysRevA.65.062305,PhysRevLett.88.187903,PhysRevA.68.052309,bellomo2017,swapping2019Castellini,saro2018,castelliniPRAmetro}, experimental evidence of spatial indistinguishability as a direct resource in quantum networks has remained challenging, because of the lack of both an informational measure of its degree and a technique to control it.

Here we experimentally implement the operational framework based on spatially localized operations and classical communication (sLOCC) which, at variance with the idea of particle locality, relies on the concept of spacial locality of measurement as in quantum field theory \cite{saro2018}. The setup is capable to control the distribution of the wave packets of two initially-uncorrelated identical photons towards two separated operational regions where the photons are collected. This allows us to continuously adjust the degree of their spatial indistinguishability at a distance, which is quantified via a suitable entropic-informational measure \cite{nosrati2019}. The two photon paths remain separated along the setup and only meet in the operational regions. By single-photon localized measurements, we prove that the two uncorrelated photons with opposite polarizations get entangled, the amount of entanglement being only related to the degree of remote spatial indistinguishability. We finally show that the nonlocal entanglement so generated is high enough to realize conditional teleportation of the state of an additional photon with fidelities higher than the classical threshold.

\begin{figure}[t!]
\centering
\includegraphics[width=0.455\textwidth]{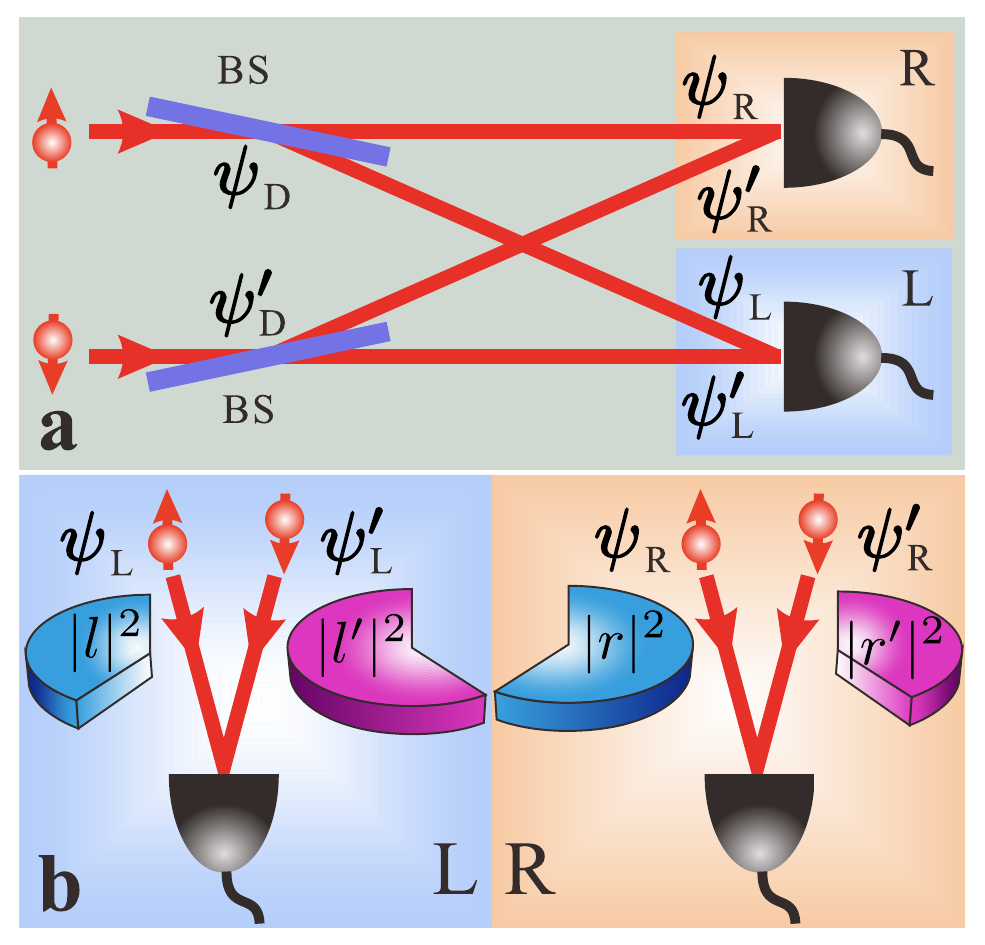}
\caption{\textbf{Theoretical scheme.} {\bf a.} The conceptual setup thought to produce and verify entanglement by sLOCC due to remote spatial indistinguishability. {\bf b.} Generic spatial distribution of particles in the state
$\ket{\Psi}=\ket{\psi_\mathrm{D} \uparrow, \psi_\mathrm{D}' \downarrow}$ before sLOCC, with $\ket{\psi_\mathrm{D}}=l \ket{\psi_\mathrm{L}}+r \ket{\psi_\mathrm{R}}$ and $\ket{\psi_\mathrm{D}'}=l' \ket{\psi_\mathrm{L}'}+r' \ket{\psi_\mathrm{R}'}$ ($0<\mathcal{I}<1$).
}\label{fig:theory}
\end{figure}

\textbf{Theory.} We start describing the basic theoretical setup thought to create and observe nonlocal entanglement by sLOCC-based indistinguishability, depicted in  Fig.~\ref{fig:theory}{\bf a}. Two separated identical particles, coming from independent sources, have pseudospins $\uparrow$ and $\downarrow$, respectively, being in the initial uncorrelated state $\ket{\psi \uparrow, \psi' \downarrow}$. Each particle wave packet is then distributed in a controllable manner towards two separated operational regions L and R by a beam splitter (BS), $\ket{\psi}\xrightarrow{\mathrm{BS}}\ket{\psi_\mathrm{D}}$ and $\ket{\psi'}\xrightarrow{\mathrm{BS}}\ket{\psi'_\mathrm{D}}$. The sLOCC measurements, represented by the two detectors in Fig.~\ref{fig:theory}{\bf a}, are realized by single-particle counting performed locally in the two regions (sLO) and coincidence measures (CC). Thinking of photons as identical particles, this scheme can be seen as a modified version of the Hanbury Brown and Twiss (HBT) experiment \cite{HBT56,HBTquanta}, the modification consisting in initially polarizing the photons and controlling the spatial distribution of their wave packets. Before sLOCC measurements, the two particles are in the state $\ket{\Psi}=\ket{\psi_\mathrm{D} \uparrow, \psi_\mathrm{D}' \downarrow}$, written in the no-label formalism \cite{LFCSciRep}, where $\ket{\psi_\mathrm{D}}=l \ket{\psi_\mathrm{L}}+r \ket{\psi_\mathrm{R}}$ ($|l|^2+|r|^2=1$) and $\ket{\psi_\mathrm{D}'}=l' \ket{\psi_\mathrm{L}'}+r' \ket{\psi_\mathrm{R}'}$ ($|l'|^2+|r'|^2=1$), with $\ket{\psi_\mathrm{X}}$, $\ket{\psi_\mathrm{X}'}$ (X $=$ L, R) indicating the two particle wave packets located in the operational region X. Notice that this state would be uncorrelated if the particles were nonidentical. In fact, independently of their spatial configuration, the particles could be individually addressed by measurements which distinguish them from one another: the state would be a product state $\ket{\psi_\mathrm{D} \uparrow}\otimes \ket{\psi_\mathrm{D}' \downarrow}$ \cite{horo2009}. Our particles are identical and not individually addressable in the measurement regions. The question then arises whether $\ket{\Psi}$ contains useful quantum correlations between pseudospins in L and R, a conceptual problem which has been longly debated \cite{saro2018}.

While particle identity is an intrinsic property of the system, one can define a continuous degree of indistinguishability which quantifies how much a given measurement process can distinguish the particles. In our setup, considering localized single-particle counting which ignores any other degree of freedom, it is natural to define a spatial indistinguishability which is nonzero when the two particles have nonzero probability to be found in both separated (remote) regions L and R. This means that the particle paths need to meet only at the detection level, so that to the eyes of the localized measurement devices the particles are indistinguishable (see Fig.~\ref{fig:theory}{\bf b}).
We use the recently-introduced entropic measure for the sLOCC-based (remote) spatial indistinguishability \cite{nosrati2019}
\begin{equation}\label{Imeas}
\mathcal{I}=-\sum_{i=1}^2 p_\mathrm{LR}^{(i)}\log_2 p_\mathrm{LR}^{(i)},
\end{equation}
where $p_\mathrm{LR}^{(1)}=|lr'|^2/(|lr'|^2+|l'r|^2)$ and $p_\mathrm{LR}^{(2)}=1-p_\mathrm{LR}^{(1)}$ (see appendix~\ref{app:A} for details). Notice that $|lr'|^2$ is the joint probability of finding a particle in L coming from $\ket{\psi_\mathrm{D}}$ and a particle in R coming from $\ket{\psi_\mathrm{D}'}$, whilst $|l'r|^2$ is clearly the vice versa (see Fig.~\ref{fig:theory}\textbf{b}). This measure ranges from
$\mathcal{I}=0$ for spatially separated wave packet distributions ($|l|^2=|r'|^2=1$ or $|l|^2=|r'|^2=0$) to $\mathcal{I}=1$ for equally distributed wave packets
($|l|^2=|l'|^2$) (see appendix~\ref{app:B}). We stress that any other degree of freedom of the particles, apart their spatial location, has no effect on $\mathcal{I}$.
Localized single-particle counting is represented by the projection operator $\hat{\Pi}_\mathrm{LR} = \sum_{\sigma,\tau = \uparrow,\downarrow}\ket{\mathrm{L}\sigma,\mathrm{R}\tau}\bra{\mathrm{L}\sigma,\mathrm{R}\tau}$ on the operational subspace, which leaves the pseudospins (and any other degree of freedom) untouched. The action of
$\hat{\Pi}_\mathrm{LR}$ on $\ket{\Psi}=\ket{\psi_\mathrm{D} \uparrow, \psi_\mathrm{D}' \downarrow}$ is predicted to produce the state \cite{saro2018}
\begin{equation}\label{entangledTheor}
|\Psi_\mathrm{LR}\rangle =(lr'|\mathrm{L} H, \mathrm{R} V \rangle
+\eta r l' |\mathrm{L} V, \mathrm{R} H \rangle)/(\sqrt{|lr'|^2+|rl'|^2}),
\end{equation}
obtained with probability $P_\mathrm{LR}=|lr'|^2+|rl'|^2$, where $\eta=\pm1$ for bosons and fermions, respectively. The amount of entanglement of this state, whose nature is intrinsically conditional, only depends on the degree of remote spatial indistinguishability $\mathcal{I}$.


\begin{figure*}
\centering
\includegraphics[width=0.85\textwidth]{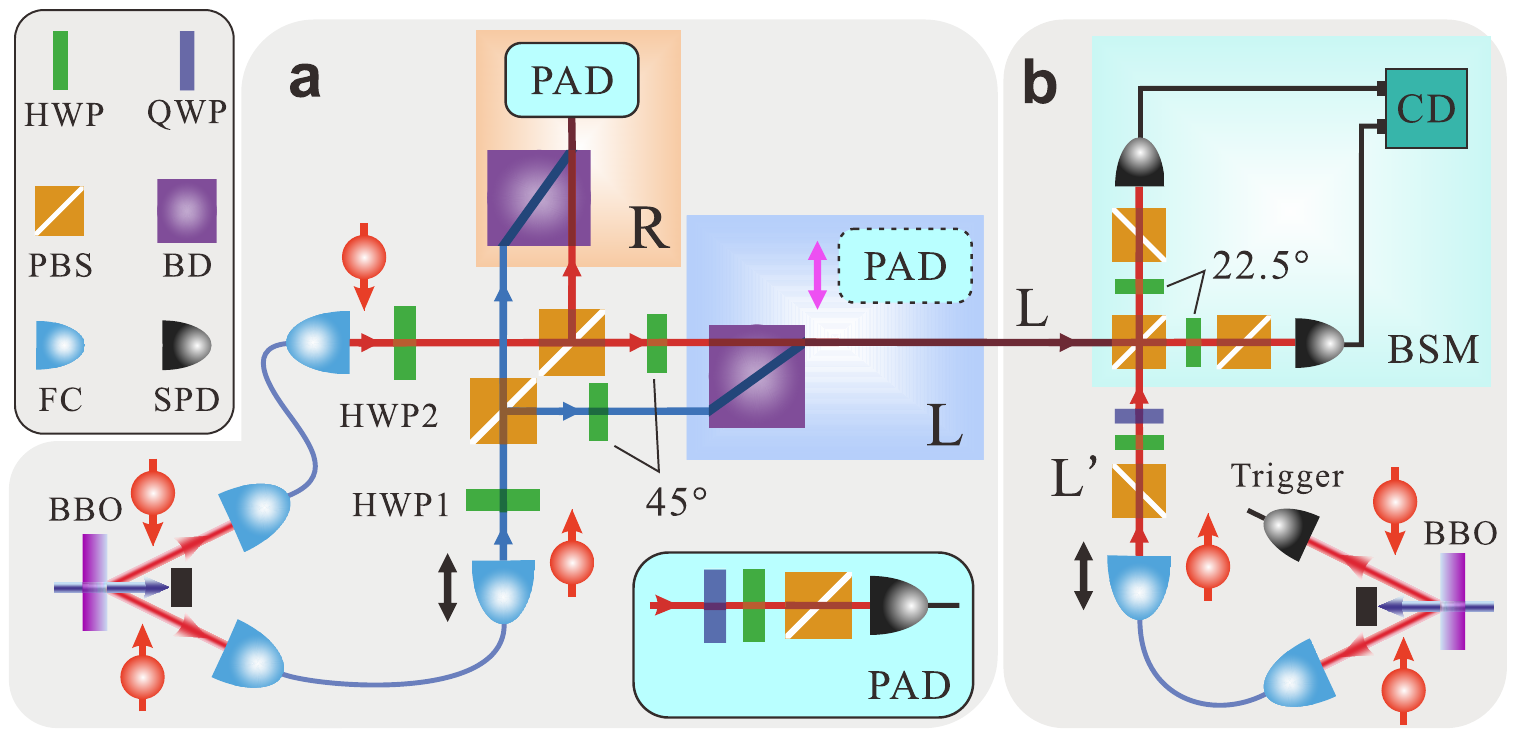}
\caption{\textbf{Experimental setup.} {\bf a.} State preparation. Independently using fiber couplers (FCs), half-wave plates (HWPs) and polarized beam splitters (PBSs), two oppositely-polarized uncorrelated photons are distributed towards two separated local regions L and R. In each region, one beam displacer (BD) is used to make the photon paths only meet at the detection level. The inset displays the unit of polarization analysis detection (PAD), including a quarter-wave plate (QWP) and a final single-photon detector (SPD). {\bf b.} Quantum teleportation realization. The PAD is removed and the photons in L proceed towards the Bell state measurement (BSM) with coincidence device (CD). The additional photon state to be teleported is generated in L$'$ (see appendix~\ref{app:C} for details).} \label{fig:setup}
\end{figure*}

\textbf{Experimental setup and preparation.}
The all-optical experimental setup which realizes the theoretical scheme above is displayed in Fig.~\ref{fig:setup}{\bf a}. Two (uncorrelated) photons are emitted via spontaneous parametric down conversion (SPDC) from a BBO crystal, designed to satisfy beamlike type-II SPDC \cite{takeuchi2001} and pumped by pulsed ultraviolet light at $400$ nm. The use of a single BBO crystal instead of two independent sources is only due to stabilization reasons. The two photons (each with wavelength $800$ nm) are initially polarized in $\ket{H}$ (horizontal, $H\equiv\uparrow$) and $\ket{V}$ (vertical, $V\equiv\downarrow$) polarization, respectively, and then collected separately by two single-mode fibers via fiber couplers (FCs). At this stage, the photons are completely uncorrelated in the product state $\ket{H}\otimes\ket{V}$, as verified with very high fidelity $(99.9 \pm 0.1)\%$ by initial measurements (see appendix~\ref{app:C}).
Before carrying on the main experiment for tuning the remote spatial indistinguishability of the two photons, we perform the usual two-photon interference to reveal the identity of the employed operational photons characterized by the visibility of Hong-Ou-Mandel (HOM) dip \cite{hom1987}, which gives a solid value of about $97.7 \%$ (see appendix~\ref{app:C}).
The BSs of the scheme in Fig.~\ref{fig:theory}{\bf a} modulating the spatial distribution of the wave packets $\psi_\mathrm{D}$, $\psi_\mathrm{D}'$ are actually substituted by the sequence of a half-wave plate (HWP$i$, $i=1,2$), a polarizing beam-splitter (PBS) and two final HWPs at $45^{\circ}$ before the location L. By rotating HWP1 and HWP2, characterized by the angles $(\pi/2-\alpha)/2$ and $-\beta/2$, respectively, we can conveniently adjust the weights of the linear spatial distribution of the photons on the two measurement locations, while the PBSs separate the different polarizations. The final HWPs at $45^{\circ}$ have the role of maintaining the initial polarization of each photon unvaried. In each region L and R one beam displacer (BD) is used to make the paths of the two photons meet at the detection level. It is straightforward to see that this setup prepares the desired state $|\Psi_\mathrm{prep}\rangle=|\psi_\mathrm{D} H, \psi_\mathrm{D}' V\rangle$ with $|\psi_\mathrm{D}\rangle=\cos\alpha\ |\psi_\mathrm{L}\rangle+\sin\alpha\ |\psi_\mathrm{R}\rangle$ and $|\psi_\mathrm{D}'\rangle=\sin\beta\ |\psi_\mathrm{L}'\rangle+\cos\beta\ |\psi_\mathrm{R}'\rangle$. By setting $\alpha$ and $\beta$, we can prepare a series of $|\Psi_\mathrm{prep}\rangle$ for different spatial distributions of the wave packets and thus of the degree of spatial indistinguishability $\mathcal{I}$. All the optical elements of the setup independently act on a single photon, so that the two single-photon states $\ket{\psi_\mathrm{D} H}$, $\ket{\psi'_\mathrm{D} V}$ are independently prepared regardless of the specific photon spatial mode (e.g., transversal electric magnetic mode such as Hermite-Gaussian mode or Laguerre-Gaussian mode).

The sLOCC measurements are implemented by single-photon detectors (SPDs) placed on both L and R for single-particle counting (sLO) and by a coincidence device (CD), which deals with the electrical signals of SPDs and outputs the coincidence counting on L and R, for classical communication (CC). An interference filter, whose full width at half maximum is 3 nm, and a single mode fiber (both of them are not shown here) are placed before each SPD. A unit of polarization analysis detection (PAD), made of a quarter-wave plate (QWP), a HWP and a PBS (see inset of Fig.~\ref{fig:setup}{\bf a}), is locally employed for verifying the predicted polarization entanglement by tomographic measurements and Bell test.

\textbf{Nonlocal entanglement from remote spatial indistinguishability.}
The setup above generates by sLOCC the distributed resource state $|\Psi_\mathrm{LR}\rangle$ of Eq.~(\ref{entangledTheor}), with $\eta=1$ (bosons), $l=\cos\alpha$, $r=\sin\alpha$, $l'=\sin\beta$ and $r'=\cos\beta$, contained in the prepared state $|\Psi_\mathrm{prep}\rangle$. As a first point, we notice that no entanglement is found when the two spatial distributions remain separated each in a local region. In fact, in this case $\mathcal{I}=0$ since the two photons are distinguished by their spatial locations. This situation is retrieved when $\alpha=\beta=0$ $(\pi/2)$, giving $|\psi_\mathrm{D}\rangle=|\psi_\mathrm{L}\rangle$ ($|\psi_\mathrm{R}\rangle$) and $|\psi_\mathrm{D}' \rangle=|\psi_\mathrm{R}'\rangle$ ($|\psi_\mathrm{L}'\rangle$). The final state of the experiment becomes the product state $|\Psi_\mathrm{LR}\rangle =\ket{\mathrm{L} H, \mathrm{R}V}$ ($\ket{\mathrm{L} V, \mathrm{R}H}$). Observations confirm this prediction of uncorrelated state with high fidelity. Notice that zero entanglement also occurs when $|\psi_\mathrm{D}\rangle$, $|\psi_\mathrm{D}'\rangle$ only share one of the two operational regions (e.g., $\alpha=\pi/4$, $\beta=0,\pi/2$), for which $\mathcal{I}=0$ (see appendix~\ref{app:D}).

\begin{figure}
\centering
\includegraphics[width=0.47\textwidth]{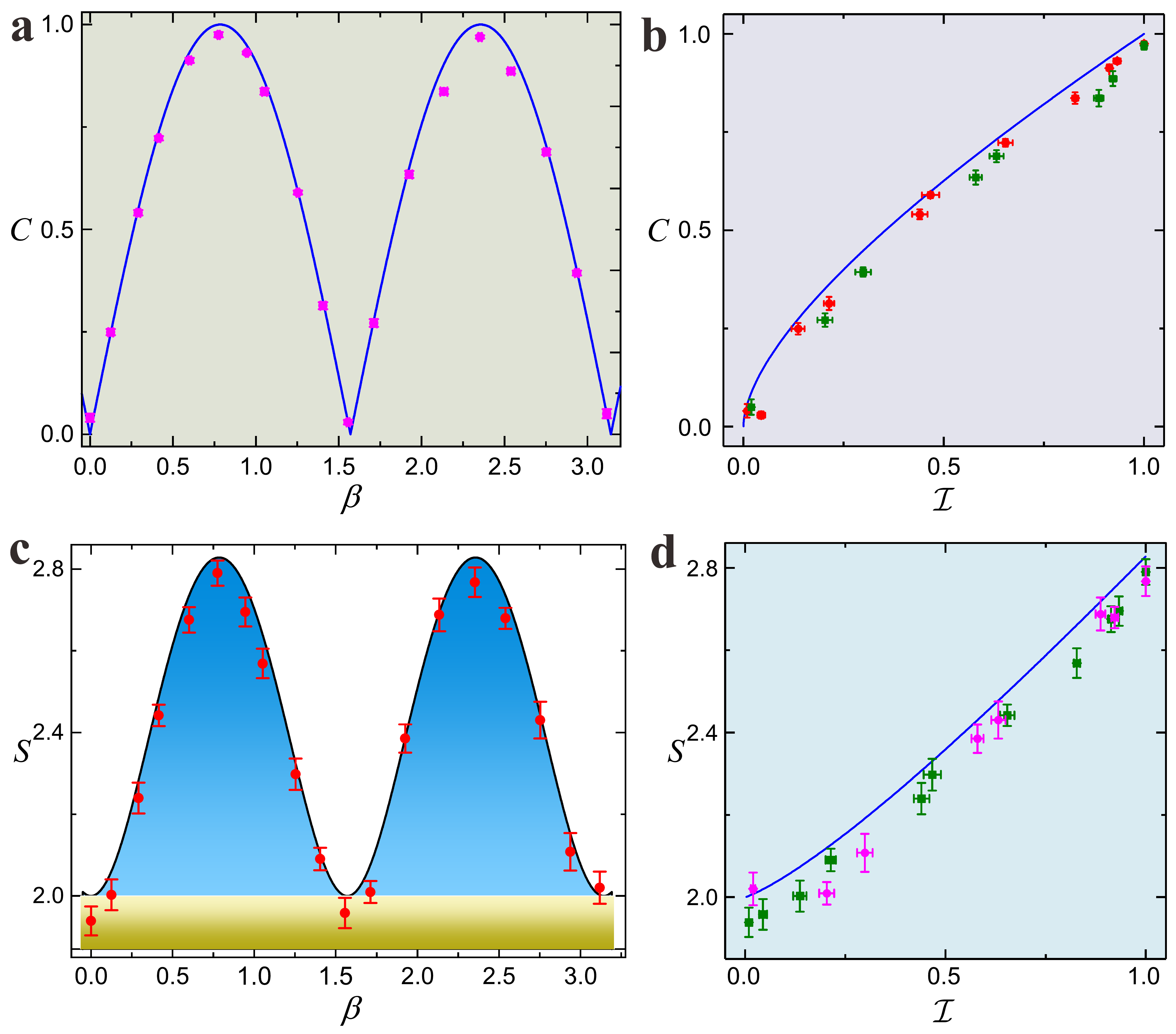}
\caption{Concurrence $C$ ({\bf a}, {\bf b}) and Bell function $S$ in the CHSH test ({\bf c}, {\bf d}), for $\alpha=\pi/4$ (parameter fixing $|\psi_\mathrm{D}$), versus $\beta$ (parameter adjusting $|\psi_\mathrm{D}'$) and $\mathcal{I}$ (remote spatial indistinguishability of Eq.~(\ref{Imeas})). Blue lines in {\bf a}, \textbf{b}, {\bf d} and black curve in {\bf c} are the theoretical predictions. Pink dots in {\bf a} and red points in {\bf c} (with error bars) are experimental results. Red and green dots in {\bf b} (pink and green dots in {\bf d}) denote experimental data for, respectively, $\beta\in[0,\pi/2]$ and $\beta\in(\pi/2,\pi]$.
}\label{fig:concur-chsh}
\end{figure}

\begin{figure}[t!]
\centering
\includegraphics[width=0.45\textwidth]{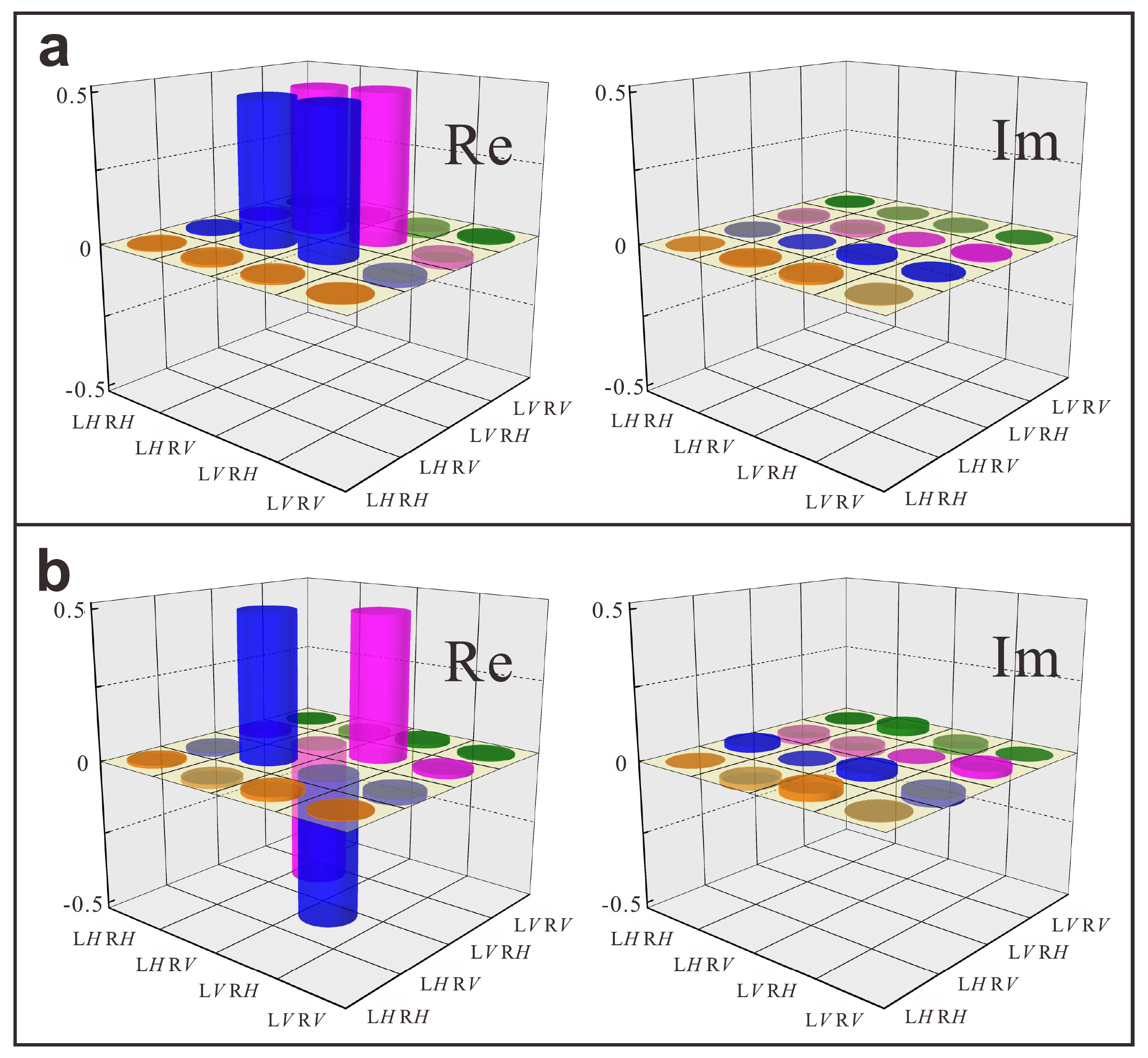}
\caption{Experimental results of the generated Bell states for $\mathcal{I}=1$. Panel {\bf a} shows the real and imaginary part of the reconstructed density matrix for $\alpha=\pi/4$ and $\beta=0.776 \pm 0.006$, the expected state being $\ket{\Psi_\mathrm{LR}^{+}}$. Panel {\bf b} shows the real and imaginary part of the reconstructed density matrix for $\alpha=\pi/4$ and $\beta=2.352 \pm 0.004$, the expected state being $\ket{\Psi_\mathrm{LR}^{-}}$.}\label{fig:entan}
\end{figure}

We verify the amount of produced entanglement as a function of $\mathcal{I}$ by adjusting different spatial distributions $|\psi_\mathrm{D}\rangle$, $|\psi_\mathrm{D}'\rangle$ on the two operational regions L and R. To this aim we fix $\alpha=\pi /4$, for which $|\psi_\mathrm{D}\rangle=
(|\psi_\mathrm{L}\rangle+ |\psi_\mathrm{R}\rangle)/\sqrt{2}$ and
\begin{equation}\label{a45}
|\Psi_\mathrm{LR}\rangle = \cos \beta\ \ket{\mathrm{L} H,\ \mathrm{R}V} + \sin\beta\ \ket{\mathrm{L} V,\ \mathrm{R}H}.
\end{equation}
The concurrence $C$ \cite{wootters1998} of this state, used to quantify the entanglement between L and R, is $C(\Psi_\mathrm{LR}) =\sin 2\beta$. From Eq.~(\ref{Imeas}), $\mathcal{I}=-\cos^2\beta\ \log_2(\cos^2\beta)-\sin^2\beta\ \log_2(\sin^2\beta)$ which coincides with the entanglement of formation $E_f(\Psi_\mathrm{LR})$ of the state $|\Psi_\mathrm{LR}\rangle$ \cite{saro2018}. Therefore, a monotonic relation $C=f(\mathcal{I})$ exists between concurrence and spatial indistinguishability of the photons. We also perform the experimental test of Bell inequality violation on the various distributed resource states $|\Psi_\mathrm{LR}\rangle$ to directly prove the presence of (nonlocal) entanglement \cite{sciarrinoPRA}. The Bell function $S$ is employed to perform a Clauser-Horne-Shimony-Holt (CHSH) test \cite{chsh1969} that, for the theoretically-predicted state of Eq.~(\ref{a45}), is $S=2\sqrt{1+C^2}=2\sqrt{1+\sin^22\beta}$ \cite{horo2009,brunner2014}. Bell inequality is violated when $S>2$, witnessing quantum correlations non-reproducible by any classical local model.
The experimental results of $C$, obtained after state reconstruction, and $S$ as functions of both $\beta$ and $\mathcal{I}$ are shown in Fig.~\ref{fig:concur-chsh}{\bf a}-{\bf d}. The results for $S$ of most of the generated states are above the classical threshold $2$ (others do not pass the CHSH test because of the imperfect behavior of experimental elements), clearly demonstrating their nonlocal entanglement.
For $\beta=\pi/4$, $|\Psi_\mathrm{LR}\rangle$ in Eq.~\ref{a45} is expected to be maximally entangled. In the experiment, in correspondence of $\beta=0.776 \pm 0.006$ we achieve $S=2.79 \pm 0.03$, which violates the Bell inequality by 26 standard deviations. The generated state holds a fidelity of $(98.6 \pm 0.2) \%$ compared to the expected Bell state $\ket{\Psi_\mathrm{LR}^{+}}$, whose reconstructed density matrix is presented in Fig.~\ref{fig:entan}{\bf a}. Moreover, when $\beta=3\pi/4$, that is when $|\psi_\mathrm{D}'\rangle=
(|\psi_\mathrm{L}'\rangle- |\psi_\mathrm{R}'\rangle)/\sqrt{2}$ and $|\psi_\mathrm{D}\rangle$ are orthogonal yet completely spatially indistinguishable ($\mathcal{I}=1$), the Bell state $\ket{\Psi_\mathrm{LR}^{-}}$ is expected. In the experiment, this state is generated with fidelity $(98.0 \pm 0.8)\%$ for $\beta=2.352 \pm 0.004$, as shown in Fig.~\ref{fig:entan}{\bf b}. Looking at the setup, these results supply strong experimental evidence that the nonlocal entanglement is activated only by remote spatial indistinguishability of photons.

To further check the correct interpretation of the findings, we also performed the measurements when the two photon paths are separated at the detection level. In this case we find no quantum correlations because, as a standard consequence of quantum mechanics, photons are distinguished and single-photon probabilities multiply (see appendix~\ref{app:D} ).

\textbf{Quantum teleportation.}
We now show that the amount of the indistinguishability-enabled entanglement is high enough to realize quantum teleportation. Notice that we can follow the steps of the standard protocol \cite{horo2009,teleportation1997,pirandola2015}, once the cases when both  photons are either in L or in R are discarded. The teleportation is here intrinsically conditional \cite{saro2018}.
The setup is presented in Fig.~\ref{fig:setup}{\bf b}. A heralded single photon is generated by pumping a second BBO crystal. One photon is used as the trigger signal while the other photon is sent to the side of L$'$ as the target to be teleported. The combination of a HWP and a QWP prepares the photon into an arbitrary state $|\phi\rangle=a\ |H\rangle+b\ |V\rangle$ with $|a|^2+|b|^2=1$. To teleport the information of $|\phi\rangle$ to the photon located in R, we perform a Bell-state measurement (BSM) between L and L$'$. Here, the Bell state $\ket{\Phi_\mathrm{LL'}}=(|\mathrm{L} H,\ \mathrm{L'} H\rangle+|\mathrm{L} V,\ \mathrm{L'} V \rangle)/\sqrt{2}$ is performed by utilizing a PBS and setting the two HWPs at $22.5^\circ$ \cite{calsa1999,calsa2001}. Correspondingly, the single-particle operation in R, rotating the state of its photon to the desired one, is $\sigma_{x}=\ket{H}\bra{V}+\ket{V}\bra{H}$, implemented by a $45^\circ$ HWP. The signals from two SPDs are processed by a CD to coincide with the signals from R and trigger. Performing quantum tomography of single-qubit state in R, we obtain the teleportation information according to four-photon coincidence counting.
Quantum teleportation is achieved exploiting the distributed Bell state
$|\Psi_\mathrm{LR}^{+}\rangle$, generated by maximum spatial indistinguishability ($\mathcal{I}=1$) of the original photons. We set $\alpha=\beta=\pi/4$, implying $|\Psi_\mathrm{LR}^{+}\rangle$ is obtained with maximum probability ($P_\mathrm{succ}=50\%$). The results of the interference in four-photon coincidence provide a visibility about $87.9\%$ (see appendix~\ref{app:E}). The eigenvectors of $\sigma_{i}$ ($i=x, y, z$) are chosen as the additional photon states $|\phi\rangle$ in L$'$ to be teleported, that is
$|\phi_\mathrm{ideal}\rangle\in \{|H\rangle, |V\rangle, \ket{\pm}=(|H\rangle\pm|V\rangle) /\sqrt{2}, \ket{\phi_\pm}=(|H\rangle\pm i |V\rangle)/\sqrt{2}\}$. First, we perform the tomography of the effectively prepared $\rho_\mathrm{prep}$ in
L$'$. Compared to $|\phi_\mathrm{ideal}\rangle$, the fidelity $F=\langle\phi_\mathrm{ideal}|\rho_\mathrm{prep}|\phi_\mathrm{ideal}\rangle$ of $\rho_\mathrm{prep}$ in our experiment is about $99.8\%$. With tomographic measurements performed in R, the experimental teleported states $\rho_\mathrm{exp}$ could be reconstructed based on the four-photon coincidence. The state fidelity $F_\mathrm{exp}$ with respect to the initially prepared state $\rho_\mathrm{prep}$ is then introduced as a figure of merit for the teleportation efficiency. For the six initial states listed above, without background subtraction, the corresponding measured fidelities are reported in 
Table~\ref{table1}, being clearly higher than the classical fidelity limit of $2/3$ \cite{popescu1995}. The quantum process matrix $\chi$ of teleportation is also reconstructed by comparing $\rho_\mathrm{prep}$ and $\rho_\mathrm{exp}$, the experimental results giving a fidelity $F_{\chi}=(79.4 \pm 1.6) \%$ (see appendix~\ref{app:E}).
All error bars in this work are estimated as the standard deviation from the statistical variation of the photon counts, which is assumed to follow a Poisson distribution.

\renewcommand\arraystretch{1.7}
\begin{table}[!]
  \centering
  \begin{tabular}{p{2cm}<{\centering}|p{1.8cm}<{\centering}|p{1.8cm}<{\centering}|p{1.8cm}<{\centering}}
  \hline
  \hline
  state & $\ket{H}$ & $\ket{V}$ & $\ket{+}$ \\ \hline
  $F_\mathrm{exp}\ (\%)$ & $90.0\pm2.0$ & $84.7\pm1.9$ & $83.1\pm2.0$  \\
  \hline
  \hline
  state & $\ket{-}$ & $\ket{\phi_-}$ & $\ket{\phi_+}$ \\ \hline
  $F_\mathrm{exp}\ (\%)$ & $82.2\pm1.9$ & $84.3\pm1.7$ & $86.3\pm2.2$ \\
  \hline
  \hline
\end{tabular}
  \caption{Fidelities of the six states $\rho_\mathrm{exp}$ with respect to the initially prepared state in the quantum teleportation process.}\label{table1}
\end{table}

\textbf{Conclusion.} We have designed a neat all-optical experiment which generates nonlocal polarization entanglement by only adjusting the degree of remote spatial indistinguishability $\mathcal{I}$ of two initially-uncorrelated photons in two separated localized regions of measurement. The value of $\mathcal{I}$ is tuned by independently controlling the spatial distribution of each photon wave packet towards the two operational regions. The photon paths only meet at the detection level. The setup implements the spatially localized (single-photon) operations and classical communication (sLOCC) \cite{saro2018} necessary to directly assess the (monotonic) relation between $\mathcal{I}$ and the amount of produced entanglement, which is verified by both state tomography and CHSH-Bell test.
We remark that the realization of the sLOCC framework excludes any possible measurement-induced entanglement to the initial state \cite{saro2018}, as may instead happen in optical interferometry with spread detection \cite{cavalcantiPRA}.
Notice that if the setup was run by two initially uncorrelated nonidentical particles, no entanglement would be in principle generated and so detected by measurements which distinguish one particle from another (LOCC).
We have finally performed teleportation between the two operational regions, with fidelities ($82$-$90$\%) above the classical threshold, by just harnessing the spatial indistinguishability of photons in those regions, with the advantage of not requiring inefficient or demanding entanglement source devices.


Our experiment physically fulfills an elementary entangling gate by simply bringing (uncorrelated) identical photons with opposite polarizations to the same operational local regions (nodes) and then accessing the (nonlocal) entanglement by sLOCC measurements. As an outlook, multiphoton entanglement can be produced by scalability of this elementary gate. 
The results ultimately prove an inherent quantum feature of composite systems: useful entanglement from controllable remote spatial indistinguishability for quantum networking. \\

\begin{acknowledgments}
This work was supported by the National Key Research and Development Program of China (Grants NO. 2016YFA0302700 and 2017YFA0304100), National Natural Science Foundation of China (Grant NO. 11821404, 11774335, 61725504, 61805227, 61805228, 61975195), Anhui Initiative in Quantum Information Technologies (Grant NO.\ AHY060300 and AHY020100), Key Research Program of Frontier Science, CAS (Grant NO.\ QYZDYSSW-SLH003), Science Foundation of the CAS (NO. ZDRW-XH-2019-1), the Fundamental Research Funds for the Central Universities (Grant NO. WK2030380017, WK2030380015 and WK2470000026).
\end{acknowledgments}


%

\newpage

\appendix

\section{Degree of remote spatial indistinguishability}\label{app:A}

In this section we summarize the theory which brings to the definition of the entropic measure of spatial indistinguishability of particles by sLOCC in two separated operational regions \cite{nosrati2019}, given in Eq.~(1) of the main text.

Consider the state of two identical two-level subsystems (particles) $\ket{\Psi^{(2)}}=\ket{\chi_1,\chi_2}=\ket{\psi_1\sigma_1,\psi_2\sigma_2}$, where $\psi_i$ ($i =1, 2$) is the spatial degree of freedom and $\sigma_i$ is the pseudospin with basis $\{\uparrow$, $\downarrow\}$). In general, $\ket{\psi_1}$ and $\ket{\psi_2}$ can be defined in the same space regions so that they spatially overlap. We are interested in defining the degree of remote spatial indistinguishability $\mathcal{I}$ of two identical particles between the local operational regions L and R, where both particles have nonzero probability to be found. Such a definition has to quantify how much the measurement process of finding one particle in each location can distinguish the particles from one another. Spatially localized operations (sLO) are thus made by single-particle local counting ignoring the pseudospins, represented by the projector
\begin{equation}\label{PiLR}
\Pi_{\mathrm{LR}}^{(2)}=\sum_{\tau_1,\tau_2=\uparrow,\downarrow}\ket{\mathrm{L}\tau_1,\mathrm{R}\tau_2}\bra{\mathrm{L}\tau_1,\mathrm{R}\tau_2},
\end{equation}
which projects the original state onto the operational subspace spanned by the computational basis $\mathcal{B}_\mathrm{LR}=\{\ket{\mathrm{L}\uparrow,\mathrm{R}\uparrow}, \ket{\mathrm{L}\uparrow,\mathrm{R}\downarrow}, \ket{\mathrm{L}\downarrow,\mathrm{R}\uparrow}, \ket{\mathrm{L}\downarrow,\mathrm{R}\downarrow}\}$. We point out that $\ket{\mathrm{L}}$ and $\ket{\mathrm{R}}$ represent, respectively, one particle in L and one particle in R. Also, classical communication (CC) is required to the detection level to know that each of the two separated regions counted one particle. Indicating with $P_{\mathrm{X\psi_i}}=|\langle \mathrm{X}|\psi_i\rangle|^2$ ($\mathrm{X}=\mathrm{L},\mathrm{R}$ and $i=1,2$) the probability of counting in $\ket{X}$ the particle coming from $\ket{\psi_i}$, the degree of the (remote) spatial indistinguishability of qubits is \cite{nosrati2019}
\begin{align}
\begin{split}
\mathcal{I}=&-\dfrac{P_{\mathrm{L\psi_1}}P_{\mathrm{R}\psi_2}}{\mathcal{Z}} \log_2 \dfrac{P_{\mathrm{L\psi_1}}P_{\mathrm{R}\psi_2}}{\mathcal{Z}}\\
&-\dfrac{P_{\mathrm{L}\psi_2}P_{\mathrm{R}\psi_1}}{\mathcal{Z}}\log_2 \dfrac{P_{\mathrm{L}\psi_2}P_{\mathrm{R}\psi_1}}{\mathcal{Z}},
\end{split}
\label{Indistinguishability}
\end{align}
where $\mathcal{Z}=P_{\mathrm{L\psi_1}}P_{\mathrm{R}\psi_2}+P_{\mathrm{L}\psi_2}P_{\mathrm{R}\psi_1}$ represent the probability that the two possible events (joint probabilities) occur.
In general, $0\leq\mathcal{I}\leq 1$. If the spatial distributions $\ket{\psi_1}$, $\ket{\psi_2}$ of the particles are separated, we have maximal information for distinguishing the particles and $\mathcal{I}=0$. Instead, $\mathcal{I}=1$ when $P_{\mathrm{L}\psi_1}=P_{\mathrm{L}\psi_2}$ and $P_{\mathrm{R}\psi_2}=P_{\mathrm{R}\psi_1}$: in this case the two spatial distributions have the same probability amplitudes (in modulus), meaning that there is no information about the origin (which way) of the two particles found one in L and one in R.

For our theoretical and experimental setup, where $\ket{\psi_1}\equiv\ket{\psi_\mathrm{D}}=l \ket{\psi_\mathrm{L}}+r \ket{\psi_\mathrm{R}}$ and $\ket{\psi_2}\equiv\ket{\psi_\mathrm{D}'}=l' \ket{\psi_\mathrm{L}'}+r' \ket{\psi_\mathrm{R}'}$, we have $P_{\mathrm{L\psi_1}} \equiv P_{\mathrm{L\psi_\mathrm{D}}}=|l|^2$,  $P_{\mathrm{R\psi_1}} \equiv P_{\mathrm{R\psi_\mathrm{D}}}=|r|^2$,   $P_{\mathrm{L\psi_2}} \equiv P_{\mathrm{L\psi_\mathrm{D}'}}=|l'|^2$, $P_{\mathrm{R\psi_2}} \equiv P_{\mathrm{R\psi_\mathrm{D}'}}=|r'|^2$. Therefore, the expression of Eq.~(\ref{Indistinguishability}) above for $\mathcal{I}$ becomes that of Eq.~(1) in the main text. Notice that this entropic measure of spatial indistinguishability at a distance is completely independent of any other degree of freedom of the particles apart their spatial location.

\begin{figure*}[t!]
\centering
\includegraphics[width=0.87\textwidth]{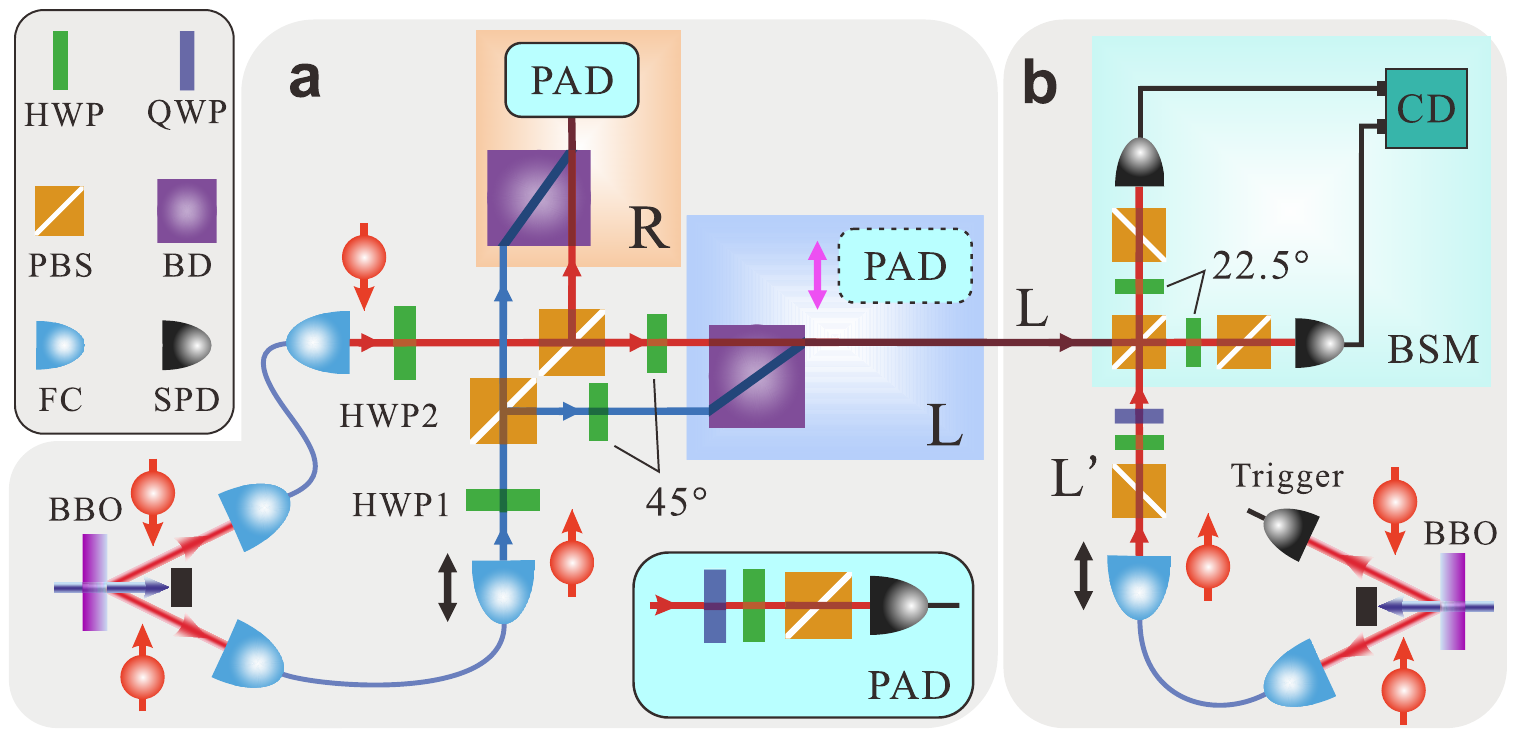}
\caption{\textbf{Sketch of the experimental setup.} {\bf a.} Two initially-independent (uncorrelated) photons from a BBO crystal with polarization $\ket{H}$ ($\uparrow$) and $\ket{V}$ ($\downarrow$), respectively, are collected separately by two single-mode fibers via fiber couplers (FCs). The control of the spatial wave function of each photon is realized by the sequence of a half-wave plate (HWP1, HWP2), a polarization beam splitter (PBS) and a final HWP at $45^\circ$ placed before L, which leaves the initial polarization of each photon unchanged. Wave packet distributions $\psi_\mathrm{D}$ and $\psi'_\mathrm{D}$, in which the photons share the same wavelength, are marked as the blue and red colors, respectively, to differentiate. In each region L and R one beam displacer (BD), made by a calcite crystal which separates the input beam into two with orthogonal polarization, is here used to make the photon paths meet at the detection level. The inset displays the unit of polarization analysis detection (PAD) at the final detection placed in L and R, consisting of a quarter-wave plate (QWP), a HWP and a PBS before the single-photon detector (SPD). A coincidence device (CD) (not shown in {\bf a}) deals with the electrical signals of the two SPDs and outputs the coincidence counting. {\bf b.} Experimental illustration of quantum teleportation. The photons in L are sent to perform the Bell state measurement (BSM) and meanwhile the PAD here which is marked by the dashed frame is removed. One of the photon pair generated by pumping another BBO is prepared as the state to be teleported, while the other photon detected directly is treated as the trigger signal. BSM is implemented with a PBS followed by two HWPs at $22.5^\circ$ and PBSs. The signals from two SPDs are processed by a CD to coincide with the signals from R and trigger.} \label{fig:setup}
\end{figure*}

\section{Spatial distribution of particle wave packets}\label{SpatialD}\label{app:B}

Consider the prepared state $|\Psi\rangle=\ket{\psi_\mathrm{D} \uparrow, \psi_\mathrm{D}' \downarrow}$, with
\begin{equation}\label{spatialstates}
\ket{\psi_\mathrm{D}}=l \ket{\psi_\mathrm{L}}+r \ket{\psi_\mathrm{R}},\quad
\ket{\psi_\mathrm{D}'}=l' \ket{\psi_\mathrm{L}'}+r' \ket{\psi_\mathrm{R}'}
\end{equation}
where $|l|^2+|r|^2=1$, $|l'|^2+|r'|^2=1$, and $\ket{\psi_\mathrm{X}}$,
$\ket{\psi_\mathrm{X}'}$ (X $=$ L, R) are the two particle wave packets located in the operational region X. With the self-evident assumption that $|\psi_\mathrm{D}\rangle$ and $|\psi_\mathrm{D}'\rangle$ cannot locate in only one measurement region simultaneously, that is $l=l'=1$ or $l=l'=0$ do not occur, we have:

\begin{itemize}
\item Zero spatial indistinguishability. This case occurs when the two wave packet distributions remain spatially separated, that is when $l=r'=1$ ($0$), giving
$|\psi_\mathrm{D}\rangle=|\psi_\mathrm{L}\rangle$ $ (|\psi_\mathrm{R}\rangle)$ and
$|\psi_\mathrm{D}'\rangle=|\psi_\mathrm{R}'\rangle$ $(|\psi_\mathrm{L}'\rangle)$.  Here the degree of remote spatial indistinguishability of Eq.~(\ref{Indistinguishability}) is clearly zero, $\mathcal{I}=0$, and the distributed state after sLOCC, coming from Eq. (2) of the main text, is the product state $|\Psi_\mathrm{LR}\rangle =\ket{\mathrm{L} \uparrow, \mathrm{R} \downarrow}$ ($\ket{\mathrm{L} \downarrow, \mathrm{R}\uparrow}$).

\item Partial spatial indistinguishability. For $0<l,  l'<1$, the spatial distributions $|\psi_\mathrm{D}\rangle$ and $|\psi_\mathrm{D}'\rangle$ are linear combinations of the particle wave packets being simultaneously in L and R with different weights. This leads to an intermediate value of the degree of spatial indistinguishability, $0<\mathcal{I}<1$, as depicted in Figure~1{\bf b} of the main text. The distributed state $|\Psi_\mathrm{LR}\rangle$ obtained by sLOCC is that of Eq.~(2) of the main text.

\item Maximum spatial indistinguishability. This case occurs when the two particle wave packets $|\psi_\mathrm{D}\rangle$ and $|\psi_\mathrm{D}'\rangle$ are equally distributed in the two operational regions, that is with equal moduli of the probability amplitudes to be located in the same regions. This means that $|l|=|l'|$ (and thus
$|r|= |r'|$). In this case the degree of remote spatial indistinguishability is maximum,
$\mathcal{I}=1$ and the distributed state of Eq.~(2) of the main text is the maximally entangled state $|\Psi_\mathrm{LR}^{\pm}\rangle =(\ket{\mathrm{L} \uparrow, \mathrm{R} \downarrow}\pm \ket{\mathrm{L} \uparrow, \mathrm{R}\downarrow})/\sqrt{2}$. A special case is retrieved when $|l=|l'|=|r|=|r'|=1/\sqrt{2}$, for which this state $|\Psi_\mathrm{LR}^{\pm}\rangle$ is generated with maximum probability ($P_{\mathrm{LR}}=50\%$) from $\ket{\Psi_{\mathrm{prep}}}$.
\end{itemize}

Summarizing, by controlling the spatial distribution of the wave packets $\ket{\psi_\mathrm{D}}$ and $\ket{\psi_\mathrm{D}'}$, we can design different configurations with various degrees of remote spatial indistinguishability, continuously ranging from $\mathcal{I}=0$ to $\mathcal{I}=1$.


\section{Experimental preparation}\label{app:C}

For the sake of convenience, we report here again in Fig.~\ref{fig:setup} the sketch of the experimental setup together with its detailed description in the caption.

\begin{figure}[t!]
\centering
\includegraphics[width=0.45\textwidth]{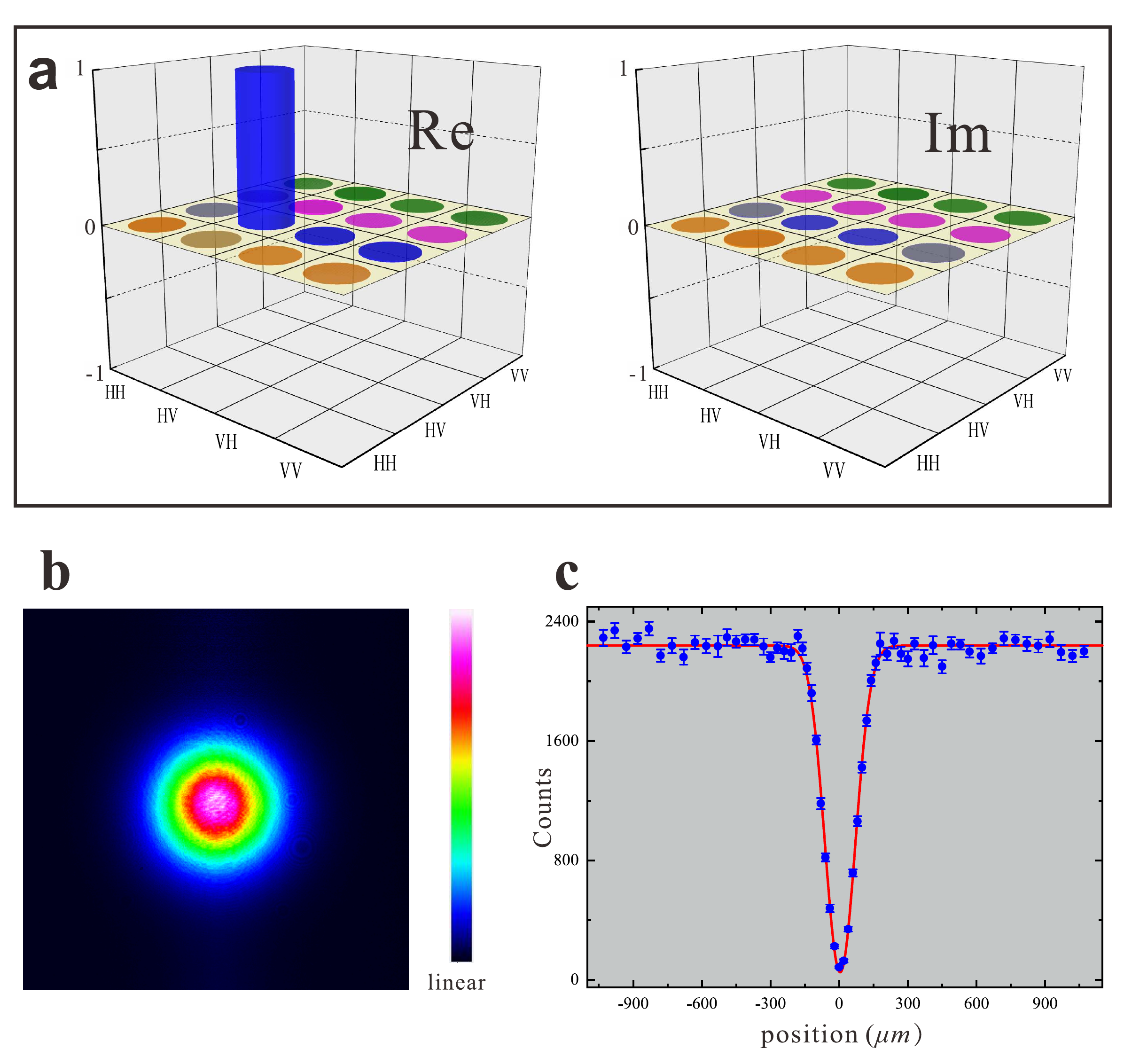}
\caption{{\bf a.} Experimental tomography of the initial state $|\psi H, \psi' V\rangle\equiv \ket{\psi H}\otimes\ket{\psi' V}$ of the two photons from BBO crystal with a high fidelity $(99.9 \pm 0.1)\%$. Two subgraphs show the real and imaginary parts of the density matrix, respectively. {\bf b.} The intensity distribution of photon with TEM (transversal electric magnetic) 00 mode generated from the single-mode fiber. {\bf c.} The HOM dip of the two-photon interference. The counts are obtained in $5$ seconds with the pumping power $30\ \mathrm{mW}$. The blue dots are the experimental results. And the red curve presents the fitting Gaussian function $G_{x}$.
}\label{initial}
\end{figure}

Two photons with the wavelength $800$nm from pumping a BBO crystal are emitted in the product state $|\Psi\rangle=|\psi H, \psi' V\rangle\equiv \ket{\psi H}\otimes\ket{\psi' V}$, where $H\equiv \uparrow$ and $V\equiv \downarrow$. The corresponding experimental reconstructed matrices of this initial state are presented in Fig.~\ref{initial}{\bf a} with a very high fidelity $(99.9 \pm 0.1)\%$. It is then verified that the two photons are initially completely uncorrelated. Each photon is then collected in a single-mode fiber via fiber couplers. In this experiment, the single-mode fiber generates an intensity shape (spatial mode) of each photon with TEM (transversal electric magnetic) 00 mode, as shown in Fig~\ref{initial}{\bf b}.

\begin{figure}[t!]
\centering
\includegraphics[width=0.45\textwidth]{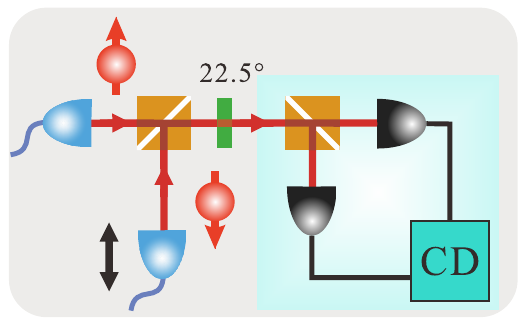}
\caption{Experimental illustration of HOM measurement. The first polarization beam splitter (PBS) combines two photons which are in the polarization $\ket{H}$ (spin up) and $\ket{V}$ (spin down), respectively. One photon is placed on a movable stage to scan the position. The photons are detected by two single photon detectors (SPD) after a half-wave plate (HWP) at $22.5^\circ$ and a second PBS. The coincidence device (CD) analyzes the signals from two SPDs and outputs the coincidence counting.
}\label{setup-hom}
\end{figure}

Before carrying on the main experiment for the continuous tuning of the degree of remote spatial indistinguishability of the two photons, we perform the usual two-photon interference and employ the visibility of Hong-Ou-Mandel (HOM) dip \cite{hom1987} to character the identity of the photons. Fig.~\ref{setup-hom} shows the setup for the HOM measurement. Fitting HOM dip data with a Gaussian function $G(x)=a\ (1-d\ e^{-b\ (x-c)^2})$, the parameters $a$, $b$, $c$, $d$ are determined. The visibility of HOM dip is defined as $\mathcal{V}=(G_\mathrm{max}-G_\mathrm{min})/G_\mathrm{max}=d$ and obtained with a value of about $97.7 \%$, which clearly reveals the identity of the employed operational photons.

A series of target states $|\Psi_\mathrm{prep}\rangle=|\psi_\mathrm{D} H, \psi_\mathrm{D}' V\rangle$ are then prepared by controlling the spatial distribution
$\ket{\psi_\mathrm{D}}$ and $\ket{\psi_\mathrm{D}'}$ of each photon wave packet towards the two separated operational regions L and R (see Fig.~\ref{fig:setup}\textbf{a}). These distributions are determined by adjusting the angles of HWP1 and HWP2. Notice that the operation of a HWP on the photon polarization with an angle $\theta$ with respect to the optical axis can be written as $\bigl( \begin{smallmatrix}
  \cos 2\theta & \sin 2\theta \\
  \sin 2\theta & -\cos 2\theta
\end{smallmatrix} \bigr)$. As shown in Fig.~\ref{fig:setup}{\bf a}, by setting HWP1 at $(\pi/2-\alpha)/2$, the photon in $\ket{\psi}$ with polarization $\ket{H}$ changes to $\sin \alpha \ket{H}+\cos\alpha \ket{V}$. Since the PBS transmits $\ket{H}$ polarization and reflects $\ket{V}$ polarization, we have $\ket{V}$ before L which should be instead $\ket{H}$ (the photon polarization has to remain invariant along the setup). Thus, before the detection device in L, a HWP at $45^\circ$ is used to flip $\ket{V}$ to
$\ket{H}$. As a result, the initial single-photon state $\ket{\psi H}$ is spatially distributed in a controlled manner by the linear combination $\ket{\psi_\mathrm{D} H}=\cos\alpha |\psi_\mathrm{L} H\rangle+\sin\alpha |\psi_\mathrm{R} H\rangle$. Following the same method with HWP2 at $-\beta/2$, it is immediate to see that the other initial single-photon state $\ket{\psi' V}$ is spatially distributed by the linear superposition
$\ket{\psi_\mathrm{D}' V}=\sin\beta |\psi'_\mathrm{L} V\rangle+\cos\beta |\psi'_\mathrm{R} V\rangle$. Since the spatial and polarization degrees of freedom are independent, the setup eventually distributes each photon wave packet towards the two operational regions
\begin{eqnarray}\label{spatialdistribution}
\ket{\psi}&\rightarrow &\ket{\psi_\mathrm{D}}=\cos\alpha |\psi_\mathrm{L}\rangle+\sin\alpha |\psi_\mathrm{R} \rangle, \nonumber\\
\ket{\psi'}&\rightarrow&\ket{\psi'_\mathrm{D}}=\sin\beta |\psi'_\mathrm{L} \rangle+\cos\beta |\psi'_\mathrm{R} \rangle.
\end{eqnarray}
These spatial states for the photons are the same of those of Eq~(\ref{spatialstates}), predicted by the theoretical scheme, with the associations $l=\cos\alpha$, $r=\sin\alpha$, $l'=\sin\beta$ and $r'=\cos\beta$.

\section{More experimental results}\label{app:D}

\begin{figure}[t!]
\centering
\includegraphics[width=0.45\textwidth]{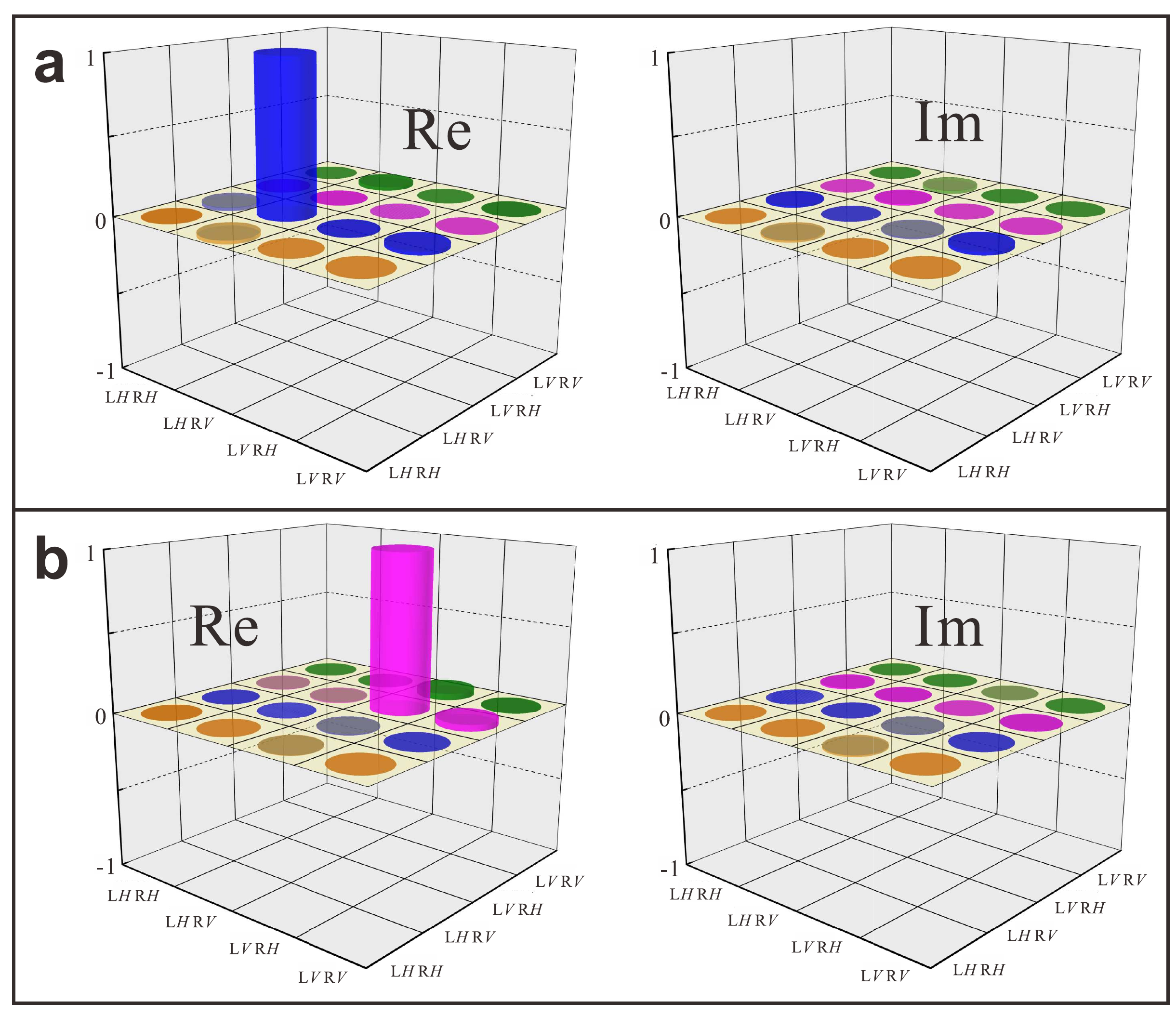}
\caption{Results for spatially separated wave packet distributions. {\bf a.} The two panels  correspond to the real and imaginary part of $|\Psi_\mathrm{LR}\rangle$, respectively, for $\alpha=\beta=0$. The fidelity is $(99.4\pm0.1)\%$. {\bf b.} The two panels  correspond to the real and imaginary part of $|\Psi_\mathrm{LR}\rangle$, respectively, for $\alpha=\beta=\pi/2$. The fidelity is $(99.8\pm0.1)\%$.
}\label{no-overlap}
\end{figure}

\begin{figure}[t!]
\centering
\includegraphics[width=0.45\textwidth]{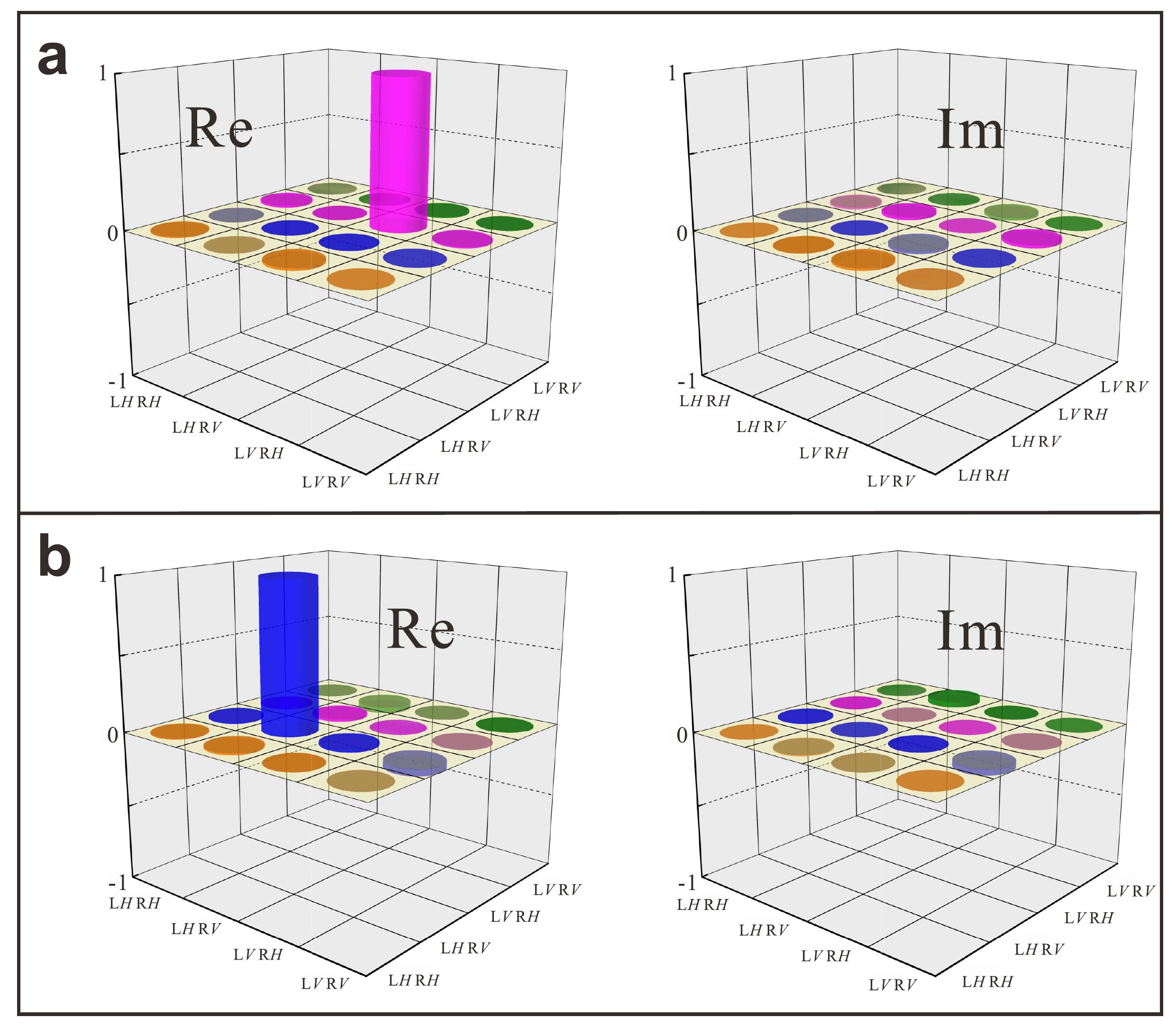}
\caption{Results for wave packet distributions sharing only one operational region. {\bf a.} Real and imaginary part of $|\Psi_{\mathrm{LR}}\rangle$ with $\alpha=\pi/4$ and $\beta=\pi/2$, with the fidelity $(99.2\pm0.1)\%$. {\bf b.} Real and imaginary part of $|\Psi_{\mathrm{LR}}\rangle$ with $\alpha=\pi/4$ and $\beta=0$ with a fidelity $(99.1\pm0.1)\%$.
}\label{par-no}
\end{figure}

As mentioned in the main text and in the section above, via changing the values of $\alpha$ and $\beta$ of HWP1 and HWP2, respectively, a series of $|\Psi_\mathrm{prep}\rangle$ with different spatial distributions of the wave packets, and thus of the degree of (remote) spatial indistinguishability $\mathcal{I}$, are prepared by the setup. The angles
$\alpha$ and $\beta$ are both fixed at the beginning of each experiment and the corresponding values are obtained from the rotated devices in which the HWPs are mounted. To describe the experimental results more accurately, the angles' values are adjusted to increase the fidelity of the experimental state compared with the desired state. In fact, as $|\Psi_\mathrm{prep}\rangle$ owns two unknown parameters, namely $\alpha$ and $\beta$, there will be several groups of solutions of $\alpha$ and $\beta$. Thus, it is necessary to fix one angle first. Here, $\alpha$ is directly read from the mounted device and the value is fixed at $\alpha=\pi/4$ in the experiment. While the other angle $\beta$ is modulated in order to increase the fidelity and the corresponding value is estimated around the display of its mounted device.

\begin{figure}[t!]
\centering
\includegraphics[width=0.47\textwidth]{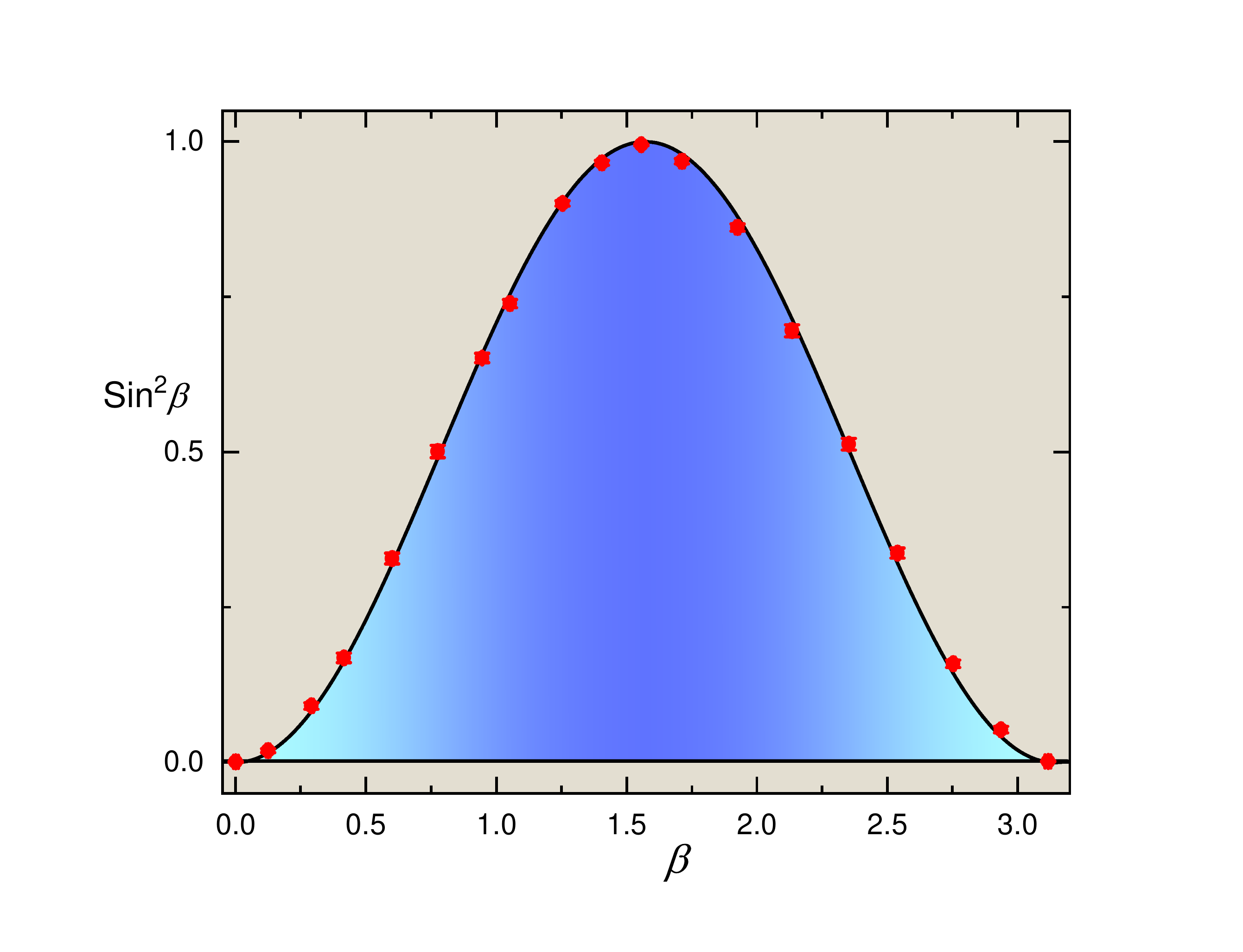}
\caption{Experimental results of probability of $|\psi'_\mathrm{D}\rangle$ in L. The black line represents theoretical prediction while the red points are the experimental results.
}\label{sin}
\end{figure}

When $\alpha=\beta=0$ ($\pi/2$), the two spatial distributions $\ket{\psi_\mathrm{D}}$ and $\ket{\psi'_\mathrm{D}}$ of Eq.~(\ref{spatialdistribution}) are separated, each one going to only one operational region. The two photons can be thus distinguished by their spatial locations, $\mathcal{I}=0$ and the distributed state $\ket{\Psi_{\mathrm{LR}}}$ obtained by sLOCC is uncorrelated in theory (see Sec.~\ref{SpatialD} above). The experimental results are presented in Fig.~\ref{no-overlap} and strongly confirm the theoretical prediction.

The condition of zero entanglement is also retrieved when the two spatial distributions of the wave packets share only one of the two operational regions. In this case, one of $\alpha$, $\beta$ is $0$ ($\pi/2$). Assuming the parameter $\alpha$ takes one of these values, one has $\sin\alpha \cos\alpha=0$ and thus $\sin\beta \cos\beta\neq0$. Fixing $\alpha=\pi/4$, this configuration is retrieved when $\beta$ is set at $0$ or $\pi/2$. The corresponding results are shown in Fig.~\ref{par-no} and confirm the theoretical predictions.

Fixing $\alpha=\pi/4$, we study the distribution of $|\psi'_\mathrm{D}\rangle$ in the operational regions L and R. Here, we choose the probability that the wave packet $|\psi'_\mathrm{D}\rangle$ is located in L to character the distribution, the theoretical prediction being $\sin^2 \beta$. The corresponding results are shown in Fig.~\ref{sin} and significantly confirm the theoretical curve.

\begin{figure}[t!]
\centering
\includegraphics[width=0.47\textwidth]{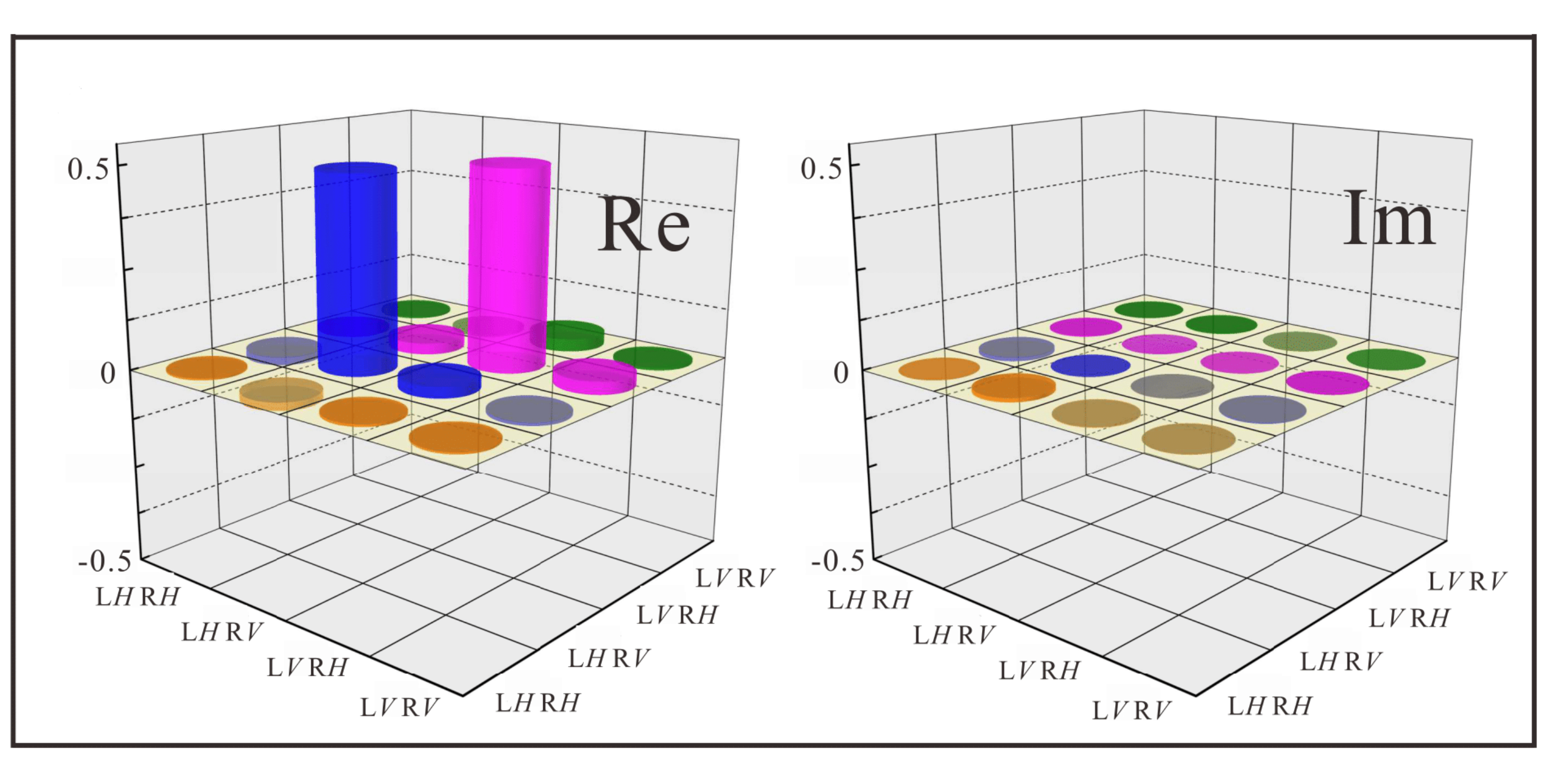}
\caption{Experimental results for separated photon paths at the detection level, showing the real and imaginary part of the distributed state.
}\label{mix}
\end{figure}

We have pointed out in the main text that the particle paths need to meet (overlap) only at the detection level, in each local region L and R, to ensure the remote spatial indistinguishability of particles. This is realized by exploiting a single mode fiber, as also mentioned in the main text. To check this, we perform other measurements for the case when the photon paths remain separated at the detection level. For this case, it is clear from standard quantum mechanics that the particles can be distinguished with the disappearance of quantum interference effect. Remote entanglement is not expected anymore. To demonstrate this other situation at the detection level, we remove the fiber coupler before HWP2 (see Fig.~\ref{fig:setup} above) transversely and replace the single mode fiber placed before single-photon detector with a multimode fiber. We set $\alpha=\beta=\pi/4$ for simplicity and the corresponding experimental results are presented in Fig.~\ref{mix}. The distributed resource state thus results to be the (unentangled) maximally mixed state $\rho_{\mathrm{LR}}=(\ket{\mathrm{L} H, \mathrm{R}V}\bra{\mathrm{L} H, \mathrm{R}V}+\ket{\mathrm{L} V, \mathrm{R}H}\bra{\mathrm{L} V,\ \mathrm{R}H})/2$. 

\begin{figure}[t!]
\centering
\includegraphics[width=0.45\textwidth]{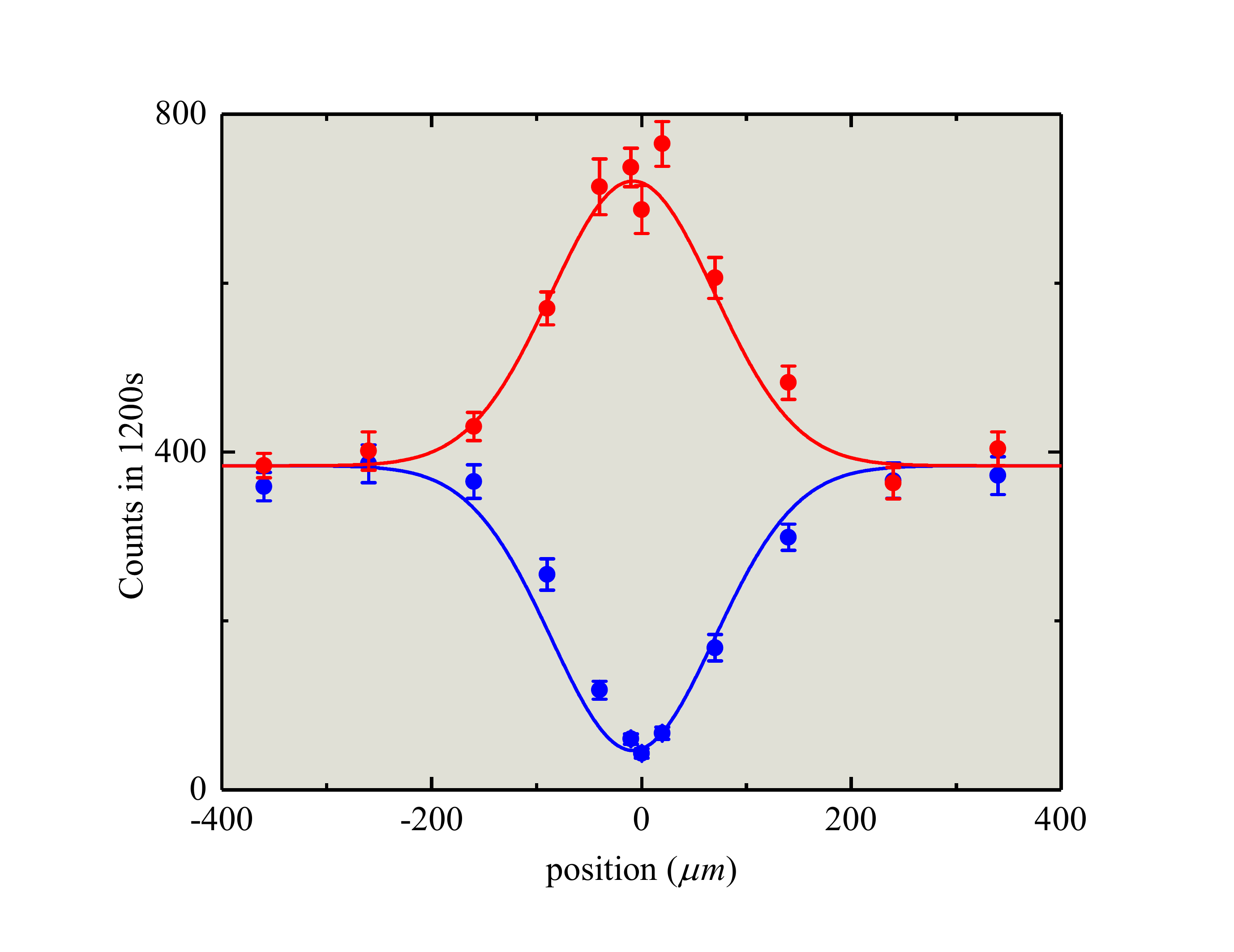}
\caption{Experimental results of temporal overlap between the photons from L and L' with the four-photon interference. The red points and curve represent the results by measuring $\ket{\Phi_{\mathrm{LL'}}^+}$ and the fitting Gaussian function, respectively. While the blue points and curve stand for the results of $\ket{\Phi_{\mathrm{LL'}}^-}$ and the corresponding fitting results, respectively. The counts are collected in $1200$ seconds with pumping power $140 \mathrm{mW}$, and shown without the background subtraction.
}\label{4photon}
\end{figure}

\begin{figure}[t!]
\centering
\includegraphics[width=0.45\textwidth]{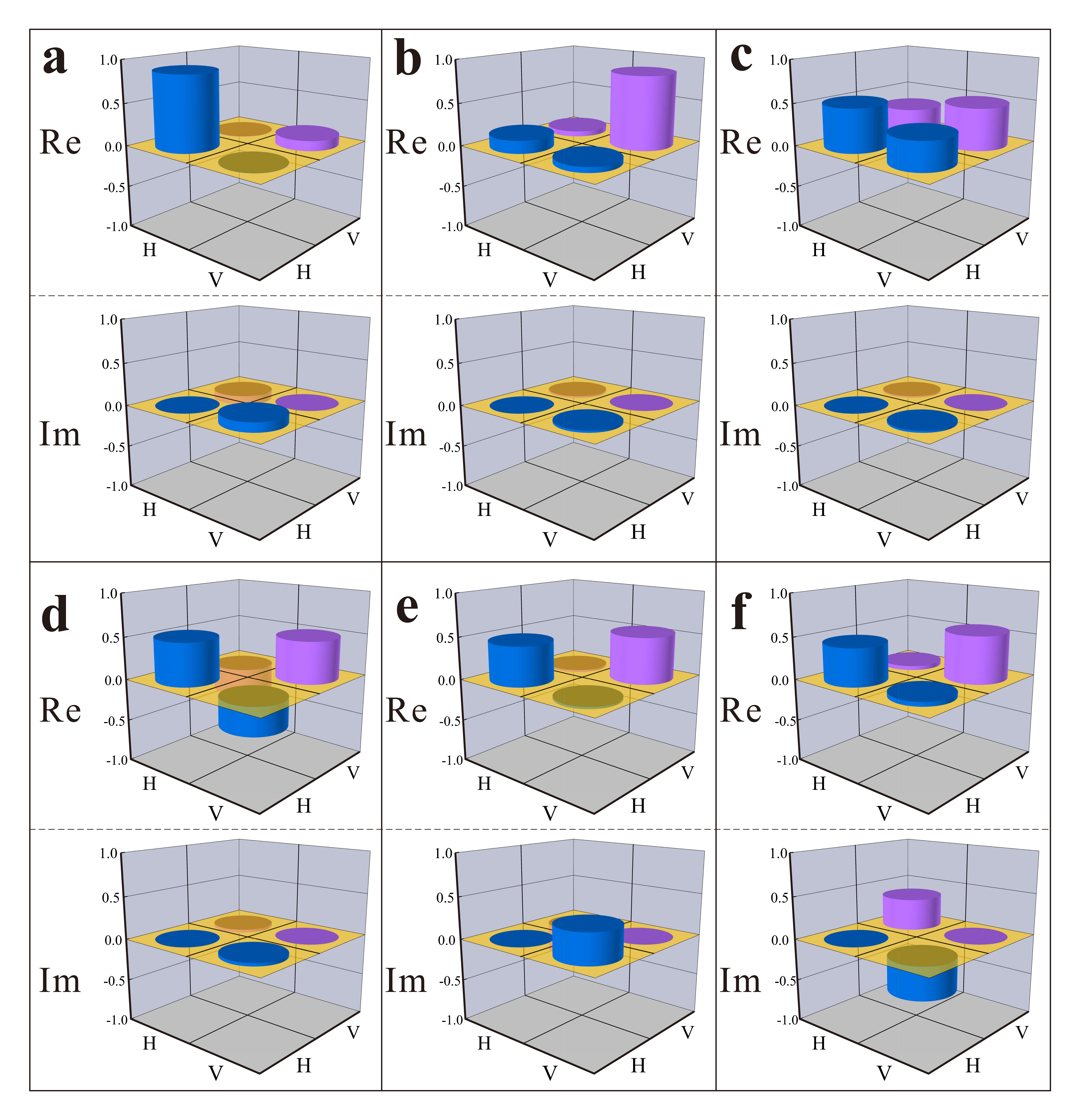}
\caption{Experimental reconstruction of the teleported states $|\phi_\mathrm{exp}\rangle$. {\bf a}, {\bf b}, {\bf c}, {\bf d}, {\bf e} and {\bf f} correspond to the density matrices of $|H\rangle$, $|V\rangle$, $\ket{+}$, $\ket{-}$, $\ket{\phi_-}$, $\ket{\phi_+}$, respectively. In each panel, the upper (lower) subgraph represents the real (imaginary) part of the density matrix.
}\label{teleportation}
\end{figure}

\begin{figure}[t!]
\centering
\includegraphics[width=0.45\textwidth]{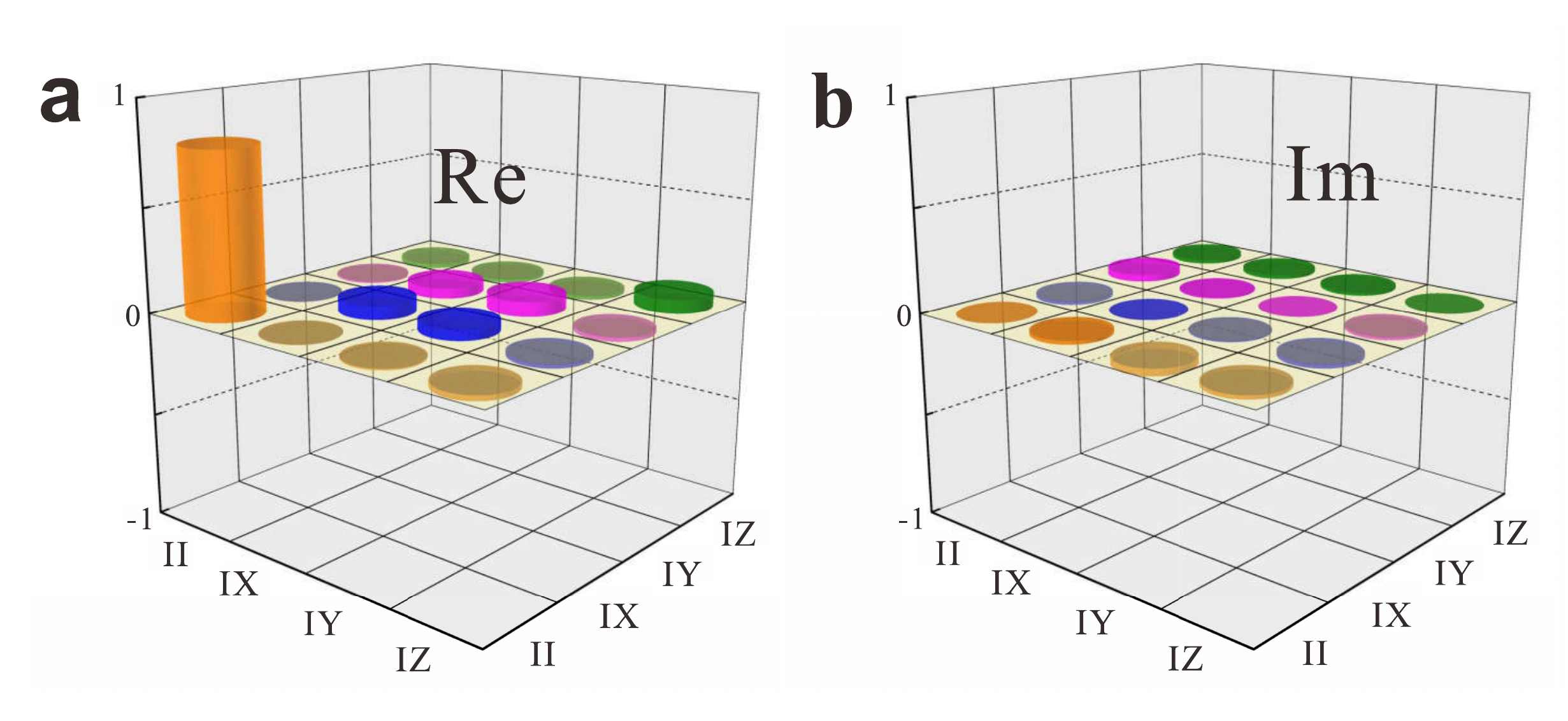}
\caption{The quantum process matrix of quantum teleportation. Experimental reconstruction of quantum process matrix $\chi$ for the teleportation protocol. {\bf a, b} show the real and imaginary parts of $\chi$, respectively. Here, $I$ is identity matrix $\sigma_1$ and $X$, $Y$, $Z$ correspond to $\sigma_{x}$, $\sigma_y$, $\sigma_z$, respectively.
}\label{result:chi}
\end{figure}

\section{More experimental results on quantum teleportation process}\label{app:E}

We investigate the temporal overlap between one of the two identical photons coming from from the operational region L and an additional (distinguishable) photon coming from L' (same lab region) in the Bell state measurement (BSM), which is verified with the interference in four-photon coincidence. As illustrated in Fig.~\ref{fig:setup}{\bf b}, by setting two HWPs at $22.5^\circ$, the measurement of the Bell state $\ket{\Phi_{\mathrm{LL'}}^+}$ is performed in BSM. If we set one of the two HWPs at $22.5^\circ$ and the other HWP at $-22.5^\circ$, the Bell state $\ket{\Phi_{\mathrm{LL'}}^-}$ is instead measured. By changing the position of the fiber coupler in L', we are able to scan the temporal overlap between L and L'. The data are obtained by measuring Bell states $\ket{\Phi_{\mathrm{LL'}}^+}$ and $\ket{\Phi_{\mathrm{LL'}}^-}$ and presented as red and blue points in Fig.~\ref{4photon}, respectively. The red points are fit with a Gaussian function $G(x)=a\ (1+d\ e^{-b\ (x-c)^2})$, while the blue ones are fit with a Gaussian function $G(x)=a\ (1-d\ e^{-b\ (x-c)^2})$. A visibility about $87.9\%$ is obtained by determining $d$.

For the six prepared states of the additional qubit to be teleported, namely $|H\rangle, |V\rangle, \ket{\pm}, \ket{\phi_\pm}$ (see main text for their expressions), the corresponding teleported state $\rho_\mathrm{exp}$ is reconstructed via performing tomographic measurements. The results are shown in Fig.~\ref{teleportation}.

To characterise quantum teleportation of the photon state from L' to R, we make the tomographic analysis of the process and reconstruct its matrix $\chi$ \cite{OBrien2004}. Here, $\chi$ is defined with $\rho_\mathrm{exp} =\Sigma_{i,\ j=1}^4\ \chi_{ij}\ \sigma_{i}\cdot \rho_\mathrm{prep}\cdot \sigma_{j}$ where $\sigma_1=I$ and $\sigma_{2, 3, 4}=\sigma_{x, y, z}$, which in the ideal case gives $\chi_{11}=1$ with all the other elements equal to $0$ \cite{chuang2010}. By assuming $\chi$ with several unknown parameters and comparing the experimental results of $\rho_\mathrm{prep}$ and $\rho_\mathrm{exp}$, we can calculate the elements of $\chi$. The experimental results are shown in Fig. \ref{result:chi} with a fidelity $F_{\chi}=(79.4 \pm 1.6) \%$.

\end{document}